\documentclass[pflatex]{tlp}

\newtheorem{theorem}{Theorem}  
\usepackage{graphicx}
 
\usepackage{url}
\usepackage{latexsym}
\usepackage{amsfonts}
\usepackage{amssymb}
 
\usepackage[pdftex]{color}

\begin{document}
\bibliographystyle{acmtrans}

\long\def\comment#1{}

\title[Querying XML Documents in Logic Programming]{Querying XML Documents \\
in Logic Programming\thanks{This work has been partially supported by the EU (FEDER) and
  the Spanish MEC under grant TIN2005-09207-C03-02.}}

\author[J. M. Almendros-Jim\'enez and A. Becerra-Ter\'on and F. J. Enciso-Ba\~nos]
{ J. M. Almendros-Jim\'enez and A. Becerra-Ter\'on and F. J. Enciso-Ba\~nos\\
 Dpto. de Lenguajes y Computaci\'on. Universidad de Almer\'\i a.\\
 \email{\{jalmen,abecerra,fjenciso\}@ual.es}
}

\submitted{2 May 2006}
  \accepted{19 October 2006}

\maketitle

\begin{abstract}
\emph{Extensible Markup Language (XML)} is a simple, very flexible text format derived from SGML. Originally designed to meet the challenges of large-scale electronic publishing, XML is also playing an increasingly important role in the exchange of a wide variety of data on the Web and elsewhere. \emph{XPath} language is the result of an effort to provide address parts of an XML document. In support of this primary purpose, it becomes in a \emph{query language} against an XML document.
In this paper we present a proposal for the implementation of the {\it XPath} language in logic programming.
With this aim we will describe the representation of XML documents by means of a \emph{logic program}. \emph{Rules} and \emph{facts} can be used for representing the document schema and the XML document itself. In particular, we will present \emph{how to index XML documents}
in logic programs: rules are supposed to be stored in \emph{main memory}, however facts are stored in \emph{secondary memory} by using two kind of indexes: one for each XML tag, and other for each group of terminal items.
In addition, we will study how to query by means of the {\it XPath} language
against a logic program representing an XML document. It evolves the \emph{spe\-cia\-li\-zation of the logic program} with regard to the {\it XPath} expression.
Finally, we will also explain how to \emph{combine the indexing and the top-down evaluation} of the logic program.
\end{abstract}


\section{Introduction}
\emph{Extensible Markup Language \emph{(XML)}} \cite{XML} is a simple, very flexible text format derived from SGML. Originally designed to meet the challenges of large-scale electronic publishing, XML is also playing an increasingly important role in the exchange of a wide variety of data on the Web and elsewhere. 

\emph{XPath} language \cite{XPath} is the result of an effort to provide address parts of an XML document. In support of this primary purpose, it becomes in a query language against an XML document, providing basic facilities for manipulation of strings, numbers and booleans. {\it XPath} uses a compact, non-XML syntax to facilitate the use of {\it XPath} within \emph{URIs} and XML attribute values. {\it XPath} operates on the abstract, logical structure of an XML document, rather than its surface syntax. {\it XPath} gets its name from its use of a path notation as in \emph{URLs} for navigating through the hierarchical structure of an XML document.

Essential to \emph{semi-structured data} \cite{AbiteboulWeb} is the selection of data
from incompletely specified data items as in an XML document. For
such data selection, the {\it XPath} language is a path language which
provides constructors similar to regular expressions and {\it ``wildcards"}
allowing a flexible node retrieval.
The \emph{XML schema}
\cite{XMLSchema}, which is also an XML document, defines the
structure of well-formed documents and thus it can be seen as a type
definition.
 
The integration of {\it logic programming languages} and \emph{web technologies}, in parti\-cular XML data processing,  
is interesting from the point of view of the applicability of logic programming.

On one hand, XML documents are the standard format of \emph{exchanging
information between applications}, therefore logic languages should be able to handle and query such
documents. 

On the other hand, logic languages could be used for \emph{extracting and inferring semantic information}
from XML documents, in the line of \emph{``Semantic Web"}  requirements \cite{Tim}. Therefore logic languages can find
a natural and interesting application field in this area.

\subsection{Contributions of this Paper}
In this paper, we are interested in the use of logic programming for handling XML documents
and {\it XPath} queries. In this context, our contributions can be summarized as follows:

\begin{enumerate}
\item An XML document can be seen as a logic program by considering \emph{facts} and \emph{rules}
 for expressing both the XML schema and document. 
 
 On one hand, rules can describe the {\it schema of
an XML document} in which a (possibly recursive) definition specifies
the well-formed documents. 

On the other hand, each XML document can be
described by means of facts, one for each terminal item (i.e. the XML tree leaves).
Although the XML schema is usually available for XML documents, our method has been studied for
extracting the XML schema
from the XML document itself. It can be consi\-dered in a certain sense as a type inference.
As future work, we will consider to adapt our technique to directly translate XML schemas into logic rules.

 \item Our second contribution is the following: once XML documents can be des\-cribed by means of a logic program, an {\it XPath}
expression against the do\-cu\-ment requires to obtain a subset of the
\emph{Herbrand model} \cite{apt} re\-presented by the logic program.
In other words, only a subset of the facts representing the XML document is required for each {\it XPath} query.

Our idea is to provide a \emph{specialization program method} in
order to retrieve only the subset of the Herbrand model required for
answering the query. In other words, we will specialize the logic
program representing an XML document with regard to an {\it XPath}
expression in order to get the answer; that is, the XML data
relevant to the query. 

Basically, the \emph{specialization
technique} will consist on \emph{specialization of rules} by
\emph{removing and reordering predicates}. It will be achieved on
the rules for the schema of the XML document, which now can be used
for retrieving a subset of the set of facts representing the XML document.
In addition, for each {\it XPath} query, a specific goal (or goals) is called, where
\emph{appropriate arguments can be instantiated}. It depends on the occurrences of boolean conditions
in the {\it XPath expression}.

\item Our technique allows the handling of XML documents as follows. 

Firstly, the XML document is loaded. It involves
the translation of the XML  document into a logic program. For efficiency reasons, the rules corresponding to the XML schema are loaded
in \emph{main memory}, but facts, which basically represent the XML document, are stored in \emph{secondary memory}
(using appropriate \emph{indexing techniques}) whenever they do not fit in main memory. 

Secondly, the user can now write queries against the loaded document. For
query solving the logic program (corresponding to the XML schema) is specialized for each query, and the top-down
evaluation of such specialized program computes the answer. The indexing technique allows that the query solving
is \emph{more efficient}, that is, it uses indexes for retrieving the facts required for the answer.

\item We have developed a prototype called {\it XIndalog} which implements {\it XPath} following the technique
presented in this paper. This prototype is hosted at \url{http://indalog.ual.es/XIndalog} 
in order to be tested. 

We have tested our prototype with not enough structured documents and complex queries, and with big documents
of different sizes. We will show benchmarks of our prototype, comparing answer times with and without our specialization
technique.

\end{enumerate}

Our approach opens two promising research lines.

\begin{itemize}

\item The first one, the extension of {\it XPath} to
a more powerful query language such as {\it XQuery} \cite{XQuery,XQuerybook,XQuery1,chamberlinXquery,wadlerEssence,wadlerAlgebra}, that is, the study of the implementation of {\it XQuery} in logic programming.

The current implementations of {\it XQuery} are implemented using as host language a functional language (see the {\it Galax} project \cite{XQuerybook,growing,simeonProjecting}). 

\item The second one, the use of logic programming as {\it inference engine} for the so-called \emph{``Semantic Web"} \cite{Tim,deckerSemanticWeb}, by introducing semantic information like \emph{RDF}  ({\it Resource Description Framework}) documents \cite{RDF} or \emph{OWL} ({\it Ontology Web Language}) specifications \cite{OWL} in the line of \cite{Wolz,Grosof,HorrocksOWL}.
\end{itemize}

\subsection{Related Work}

The integration of {\it declarative programming} and {\it XML data processing} is a research field of increasing interest
in the last years. There are proposals of new languages for XML data processing based 
on functional, and logic programming (see \cite{survey} for a survey). In addition, {\it XPath} and {\it XQuery} have been also implemented in declarative languages. 

The most relevant contribution is the {\it Galax} project \cite{simeonProjecting,XQuerybook}, which is an implementation of {\it XQuery} in functional programming, using {\it OCAML}\cite{ocaml} as host language. There are also proposals for new languages based on functional programming rather than implementing {\it XPath} and {\it XQuery}.
This is the case of {\it XDuce} \cite{hosoyaXduce} and {\it CDuce} \cite{CDuce}, which are 
languages for XML data processing, using regular expression pattern matching over XML trees, subtyping as basic mechanism, and {\it OCAML} as host language. The {\it CDuce} language does fully statically-typed transformation of XML documents, thus guaranteeing correctness. In addition, there are proposals around {\it Haskell} for the handling of XML documents, such as {\it HaXML} \cite{HaXML,UUXML} and \cite{Wallace}.

There are also contributions in the field of logic programming for the handling of XML documents.
For instance, the {\it Xcerpt} project \cite{Xcerpt,xcerptnew} proposes a pattern and rule-based query language for XML documents, using the so-called query terms including
logic variables for the retrieval of XML elements. For this new language a specialized unification algorithm
for query terms has been studied in \cite{Simulation}. Another contribution of a new language is {\it XPathLog} (the {\it Lopix } system)
\cite{XPathLog} which is a {\it Datalog}-style extension for {\it XPath} with variable bindings.
{\it Elog} \cite{elog} is also a logic-based XML data manipulation language, which has been used for representing Web documents by means of logic programming.
This is also the case of {\it XCentric} \cite{typeXML,flex}, which can represent XML documents by means of logic programming, and
handles XML documents by considering terms with functions of flexible arity and regular types.
Finally, {\it FNPath} \cite{fnpath} is a proposal in order to use {\it Prolog} as query language 
for XML documents based on a field-notation, for evaluating {\it XPath} expressions based on {\it DOM}.

The {\it Rule Markup Language} ({\it RuleML}) \cite{RuleML,RuleML2,boleyRDF} is a different kind of proposal in
this research area. The aim of {\it RuleML} is the representation of {\it Prolog} facts and rules in
XML documents, and thus, the introduction of {\it rule systems} into the {\it Web}. 

Finally, some well-known {\it Prolog} implementations include libraries for loading 
and querying XML documents, such as {\it SWI-Prolog} \cite{swi} and {\it CIAO} \cite{pillow}.

In the cited logic approaches interested in {\it XPath} queries
\cite{Xcerpt,XPathLog} {\it XPath} is directly handled, that is, rules and
queries use a new kind of {\it Prolog} terms adapted to XML patterns.
It involves to study new unification algorithms for the new {\it Prolog} terms. However,
in our work we will show how to handle XML documents not introducing new {\it Prolog} terms, but using
the standard {\it Prolog} terms. In addition, in our case, {\it XPath} queries evolve a
program transformation.  The top-down evaluation of the goals w.r.t. the transformed program obtains a set of answers which represents a subset of the Herbrand model of the transformed program.
This subset allows the reconstruction of the XML document representing the answer.
The reconstruction follows the same criteria as the translation of XML document-logic program.

Our proposal requires the representation of XML documents into logic programming, and thus 
it can be  compared with those ones representing XML documents in 
logic programming (for instance, \cite{Xcerpt,typeXML,pillow,swi}) and, with those ones representing XML documents in relational databases (for instance, \cite{monetdb,ordpath,dewey}). 
In our case, rules are used for expressing the structure of well-formed XML documents, and XML elements are represented by means of facts. Moreover, our handling of XML documents is more {\it ``database-oriented"} since we use secondary memory and file indexing in order to   retrieve the database records. The reason for such decision is that XML documents can usually be too big for main memory \cite{simeonProjecting}.

With regard to {\it RuleML} \cite{RuleML}, we translate XML documents into a logic program using facts
and rules; however we are not still interested in the
translation of logic rules into XML (or RDF) documents. This translation would be interesting when
semantic information is handled by means of logic programming. In fact, our idea is to consider these aspects as future work in the line of \cite{Wolz,Grosof,HorrocksOWL}.

There is an analogy among our specialization technique and the {\it
magic sets}-based program specialization technique  used for
deductive databases, which uses the {\it bottom-up} evaluation for
answering queries. We have also studied such  technique for XML
documents in a previous work \cite{jesusjucs}. In fact, we have
developed two releases of {\it XIndalog}: one of them implements the
top-down approach presented in this paper and the other one
implements the bottom-up approach. 

The main differences between the
top-down and the bottom-up approaches are the program transformation technique and evaluation method of
queries. 
In the second case, we use: (1) the \emph{fix-point} operator in order to evaluate {\it XPath}
queries, and (2) a \emph{magic sets} based technique in order to specialize and evaluate the
program. With respect to the transformation of XML documents into a logic program, let us
remark that this one in both approaches is the same. However, the specialization technique is
different, the technique of this paper is based on predicate removing and
reordering, and the instantiation of the goals called in a top-down fashion.

\subsection{Structure of the Paper}

The structure of the paper is as follows. Section 2 will review basic concepts of XML documents and {\it XPath} queries.
Section 3 will study the translation of
XML documents into {\it Prolog}; section 4 will present  the program specialization technique applied to {\it XPath} queries;
section 5 will prove theoretical results about our technique;
 section 6 will show the indexing technique over XML documents represented by means
of logic programming and will explain the combination of the indexing and program specialization techniques;
section 7 will show
 the Web prototype developed under {\it SWI-Prolog} for the language
{\it XPath} at the University of Almeria
(\url{http://indalog.ual.es/Xindalog}), presenting benchmarks of our prototype;
and finally, section 8 will conclude and present future work.

\section{XML and XPath}

An \emph{XML document} basically is a \emph{labeled tree} with inner nodes
representing \emph{composed or non-terminal items} and leaves
representing \emph{values or terminal items}.
For instance, let us consider
the following XML document which we will use in the paper as running example:
\begin{center}
{\fontsize{7pt}{9pt}\selectfont
{\it
\noindent \begin{tabular}{l}
\hline
$<$books$>$\\
$<$book year=$``$2003$"$$>$\\
$<$author$>$Abiteboul$<$/author$>$\\
$<$author$>$Buneman$<$/author$>$\\
$<$author$>$Suciu$<$/author$>$\\
$<$title$>$Data on the Web$<$/title$>$\\
$<$review$>$A $<$em$>$fine$<$/em$>$ book.$<$/review$>$\\
$<$/book$>$\\
$<$book year=$``$2002$"$$>$\\
$<$author$>$Buneman$<$/author$>$\\
$<$title$>$XML in Scotland$<$/title$>$\\
$<$review$>$$<$em$>$The $<$em$>$best$<$/em$>$ ever!$<$/em$>$$<$/review$>$\\
$<$/book$>$\\
$<$/books$>$\\
\hline
\end{tabular}
}
}
\end{center}
\noindent In the XML document, the
tags are used for specifying  a set of $\it books$ described by means of author's names, the title and a review. Each $\it book$ is qualified by means an attribute called $\it year$.
For each element $\it book$, we have three grouped subelements $\it author$,
$\it title$ and $\it review$. In addition, the element $\it review$ contains subelements used for formatting
 the text described by the review.

Here, the XML database includes two books. The first one, edited in {\it 2003}, with {\it author}s
 {\it Abiteboul}, {\it Buneman} and {\it Suciu}, and {\it title} {\it``Data on the Web"}.
 Finally, the opinion of the reviewer for this book was:
  {\it ``A} {\em fine} {\it book"}. The second one, edited in {\it 2002}, was written by
 {\it Buneman} with title {\it XML in Scotland}, and the opinion of the reviewer was
 {\em ``The best ever!"}.

 XML documents describe data by means of a 
 \emph{semi-structured data model} \cite{AbiteboulWeb}, whose main features are the
 occurrences of \emph{heterogeneous records}, and in particular, \emph{non-first normal relations},
\emph{missing values}, among others.

Now, with respect to the above XML document,
 we can consider the following two {\it XPath} expressions, as well as
 the expected answers in XML format:
\begin{center}
{\fontsize{7pt}{9pt}\selectfont
{\it
\noindent \begin{tabular}{cc}
{\it XPath Expression} & \hspace*{-3cm}{\it Expected XML Answer}\\
\hline
{\bf (1)} /books/book[author=``Suciu"]/title & \hspace*{-3cm}{\bf (1)} $<$title$>$Data on the Web$<$/title$>$\\
---------------------------------------- & \hspace*{-3cm} ----------------------------------------\\
{\bf (2)} /books//title &
\hspace*{-3cm}\begin{tabular}{l}
{\bf (2)} $<$title$>$Data on the Web$<$/title$>$\\
{\bf (2)} $<$title$>$XML in Scotland$<$/title$>$
\end{tabular}\\
\hline
\end{tabular}
}
}
\end{center}

\noindent where {\bf (1)} requests {\it Suciu}'s book titles, and {\bf (2)} requests book titles
without taking into account the structure of the book records.

\section{Translating XML Documents into Logic Programming}
In this section, we will show how to translate an XML document into a logic program.
We will use a set of rules for describing the XML schema and a set of facts for storing the XML document.

In general, an XML document includes
\begin{itemize}
\item[(a)] \emph{tagged elements} which have
the form:
\begin{center}
$\it <tag~att_1=v_1,\dots,att_n=v_n > subelem_1,\dots,subelem_k </ tag>$
\end{center}
\noindent where $\it att_1,\dots,att_n$ are the attributes names, $\it v_1,\dots,v_n$ are the attribute values supposed
to have a {\it basic type}: strings, integers, real numbers, lists of integers or real numbers, 
and $\it subelem_1,\dots,subelem_k$ are subelements; and
\item[(b)] \emph{untagged elements} which have a basic type.
\end{itemize}
 
\emph{Terminal tagged elements} (i.e. XML tree leaves) are those ones whose subelements have a basic type
and do not have attributes. Otherwise they are called \emph{non-terminal tagged
elements} (i.e. inner nodes).
Two tagged elements are \emph{similar} whether they have the same structure;
that is, they have the same tag
and attributes names, and the subelements are similar.
Untagged elements are always similar.
Two tagged elements are \emph{distinct} if they do not have the same tag and, finally,
they are \emph{weakly distinct} if they have the same tag but they are not similar.

\subsection{Numbering XML documents}

In order to define our translation we need to number the nodes of the XML document.
Similar kinds of node numbering have been studied in some works about XML processing
in relational databases \cite{monetdb,ordpath,dewey}. Our goal is similar to these approaches:
to identify each inner node and leaf of the tree represented by the XML document.

Given an XML document we can consider a new XML document called \emph{node-numbered XML document} as follows. Starting
from the root element numbered as $1$, the node-numbered XML document is numbered using an attribute called {\bf nodenumber}\footnote{It is supposed that ``nodenumber" is not already used as attribute in the original XML document.} where
each $\it j$-th child of a tagged element is numbered with the sequence of natural numbers $\it i_1.\dots.i_t.j$ whenever
the parent is numbered as $\it i_1.\dots.i_t$:
\begin{center}
\begin{tabular}{l}
$\it <tag~att_1=v_1,\dots,att_n=v_n,{\bf nodenumber=i_1.\dots.i_t.j}>$\\
\hspace*{2cm}$\it elem_1,\dots,elem_s</ tag>$\\
\end{tabular}
\end{center}
This is the case of tagged elements; If the $j$-th child has a basic type and the parent is a non-terminal
tagged element then the element is labeled and numbered as follows:
$$\it <unlabeled~{\bf nodenumber=i_1.\dots.i_t.j}> elem </ unlabeled>$$ 
Otherwise the element is not numbered.
It gives to us a \emph{hierarchical and left-to-right numbering} of the nodes of an
XML document.  An element in an XML document is further left in the XML
tree than another when the node number is smaller w.r.t. the
lexicographic order on sequences of natural numbers. The node
numbered XML document corresponding to the running example is as
follows:
\begin{center}
{\fontsize{7pt}{9pt}\selectfont
{\it
\noindent \begin{tabular}{l}
\hline
$<$books nodenumber=1$>$\\
$<$book year=$``$2003$"$, nodenumber=1.1$>$\\
$<$author nodenumber=1.1.1$>$Abiteboul$<$/author$>$\\
$<$author nodenumber=1.1.2$>$Buneman$<$/author$>$\\
$<$author nodenumber=1.1.3$>$Suciu$<$/author$>$\\
$<$title nodenumber=1.1.4$>$Data on the Web$<$/title$>$\\
$<$review nodenumber=1.1.5$>$\\
$<$unlabeled nodenumber=1.1.5.1$>$ A $<$/ unlabeled$>$\\
$<$em nodenumber=1.1.5.2$>$fine$<$/em$>$\\
$<$unlabeled nodenumber=1.1.5.3$>$ book.  $<$/ unlabeled$>$\\
$<$/review$>$\\
$<$/book$>$\\
$<$book year=$``$2002$"$ nodenumber=1.2$>$\\
$<$author nodenumber=1.2.1$>$Buneman$<$/author$>$\\
$<$title nodenumber=1.2.2 $>$XML in Scotland$<$/title$>$\\
$<$review nodenumber=1.2.3 $>$\\
$<$em nodenumber=1.2.3.1$>$\\
$<$unlabeled nodenumber=1.2.3.1.1$>$ The $<$/unlabeled$>$\\
$<$em nodenumber=1.2.3.1.2$>$best$<$/em$>$\\
$<$unlabeled nodenumber=1.2.3.1.3$>$ ever! $<$/unlabeled$>$\\
$<$/em$>$\\
$<$/review$>$\\
$<$/book$>$\\
$<$/books$>$\\
\hline
\end{tabular}
}
}
\end{center}
In addition, we have to consider a new document called \emph{type and node-numbered XML document} numbered using an attribute called {\bf typenumber} as follows. Starting the numbering from $1$ in the root of the node-numbered XML document, each tagged element is numbered as:
\begin{center}
\begin{tabular}{l}
$\it <tag~att_1=v_1,\dots,att_n=v_n,nodenumber=i_1.\dots,i_t.j,{\bf typenumber=k}>$\\
\hspace*{2cm}$\it elem_1,\dots,elem_s </ tag>$\\
\end{tabular}
\end{center}
and
\begin{center}
\begin{tabular}{l}
$\it <unlabeled~nodenumber=i_1.\dots.i_t.j,{\bf typenumber=k}>$\\
\hspace*{2cm}$\it elem</ unlabeled>$
\end{tabular}
\end{center}
for ``unlabeled'' nodes. In both cases, the type number of the tag
is $k=l+n+1$ whenever the type number of the parent is $l$, and
$n$ is the number of tagged elements weakly distinct to the
parent, occurring in leftmost positions at the same level of the
XML tree. Therefore, all the children of a tag have the same type
number.

\begin{figure*}[!t]
\begin{center}
\fbox{
\includegraphics[width=12cm,height=10cm]{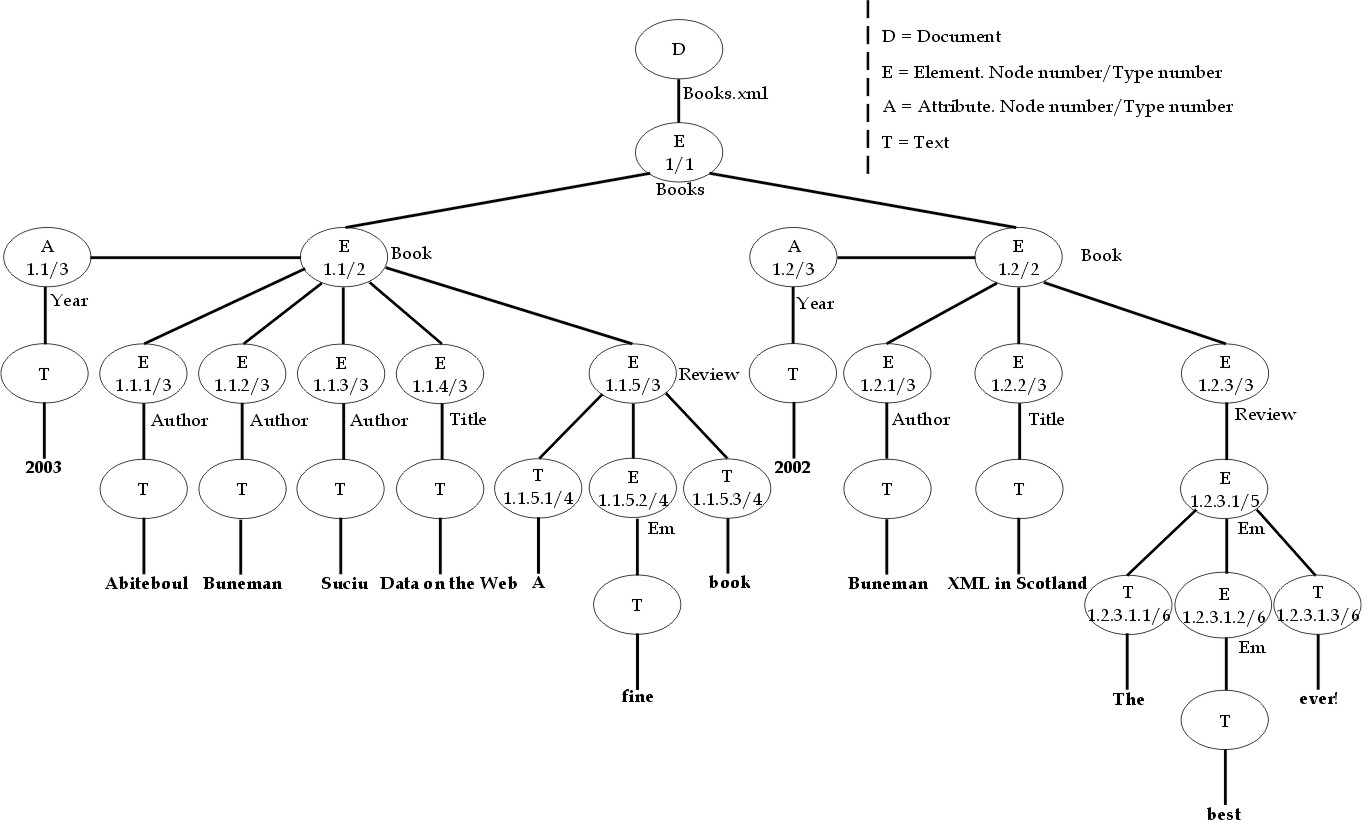}
}
\caption{{\bf Type and node numbering in the XML tree of the running example}}
\label{prole2005}
\end{center}
\end{figure*}

For instance, with respect to the running example, we can see in the Figure \ref{prole2005}
the type and node numbering which represent the following type and node numbered XML document.

\begin{center}
{\fontsize{7pt}{9pt}\selectfont
{\it
\noindent \begin{tabular}{l}
\hline
$<$books nodenumber=1, typenumber=1$>$\\
$<$book year=$``$2003$"$, nodenumber=1.1, typenumber=2$>$\\
$<$author nodenumber=1.1.1 typenumber=3$>$Abiteboul$<$/author$>$\\
$<$author nodenumber=1.1.2 typenumber=3$>$Buneman$<$/author$>$\\
$<$author nodenumber=1.1.3 typenumber=3$>$Suciu$<$/author$>$\\
$<$title nodenumber=1.1.4 typenumber=3$>$Data on the Web$<$/title$>$\\
$<$review nodenumber=1.1.5 typenumber=3$>$\\
$<$unlabeled nodenumber=1.1.5.1 typenumber=4$>$ A $<$/ unlabeled$>$\\
$<$em nodenumber=1.1.5.2 typenumber=4$>$fine$<$/em$>$\\
$<$unlabeled nodenumber=1.1.5.3 typenumber=4$>$ book.  $<$/ unlabeled$>$\\
$<$/review$>$\\
$<$/book$>$\\
$<$book year=$``$2002$"$ nodenumber=1.2, typenumber=2$>$\\
$<$author nodenumber=1.2.1 typenumber=3$>$Buneman$<$/author$>$\\
$<$title nodenumber=1.2.2 typenumber=3$>$XML in Scotland$<$/title$>$\\
$<$review nodenumber=1.2.3 typenumber=3$>$\\
$<$em nodenumber=1.2.3.1 typenumber=5$>$\\
$<$unlabeled nodenumber=1.2.3.1.1, typenumber=6$>$ The $<$/unlabeled$>$\\
\hline
\end{tabular}
}
}
\end{center}

\begin{center}
{\fontsize{7pt}{9pt}\selectfont
{\it
\noindent \begin{tabular}{l}
\hline

$<$em nodenumber=1.2.3.1.2, typenumber=6$>$best$<$/em$>$\\
$<$unlabeled nodenumber=1.2.3.1.3,typenumber=6$>$ ever! $<$/unlabeled$>$\\
$<$/em$>$
$<$/review$>$\\
$<$/book$>$\\
$<$/books$>$\\
\hline
\end{tabular}
}
}
\end{center}

Let us focus our attention to the type numbering of {\it review}. According
to the proposed type numbering, the children of {\it review} are numbered
as $k=l+n+1$ where $l$ is the type number of {\it review}, and
$n$ is the number of {\it weakly distinct records} of {\it  review} at the same
level of the tree. Therefore, the first set of children is numbered as 
$4=3+0+1$ and the second set of children is numbered as $5=3+1+1$
(i.e. the first and second reviews are weakly distinct).
This kind of type numbering allows us to distinguish both kind of
records and not to confuse them.

Let us remark that in practice the type and node numbering of XML documents can be simultaneously
generated at the same time as the translation into the logic program. In fact, the type and node numbered
version of the original XML document is not generated as an XML file.

\subsection{Translation of XML documents}

Now, the translation of the XML document into a logic program ${\cal P}$ is as follows.
For each non-terminal tagged element in the type and node numbered XML document:
\begin{center}
\begin{tabular}{l}
$\it <tag~att_1=v_1,\dots,att_n=v_n,nodenumber=i,typenumber=k>$\\
\hspace*{2cm}$\it elem_1,\dots,elem_s </ tag>$\\
\end{tabular}
\end{center}
we consider the following rule, called \emph{schema rule}:

\noindent \begin{tabular}{ll}
\hline
$\it tag(tagtype(Tag_{i_1},\dots,Tag_{i_t},$ & $\it [Att_1,\dots,Att_n]),NodeTag,k)$:-\\
& \hspace*{-3,5mm}$\it tag_{i_1}(Tag_{i_1},[NodeTag_{i_1}|NodeTag],r)$,\\
& \hspace*{-3,5mm}$\dots$,\\
& \hspace*{-3,5mm}$\it tag_{i_t}(Tag_{i_t},[NodeTag_{i_t}|NodeTag],r)$,\\
& \hspace*{-3,5mm}$\it att_1(Att_1,NodeTag,r)$,\\
& \hspace*{-3,5mm}$\dots$,\\
& \hspace*{-3,5mm}$\it att_n(Att_n,NodeTag,r).$\\
\hline
\end{tabular}
where
\begin{itemize}
\item {\it tagtype} is a new function symbol used for building a {\it Prolog} term containing the XML document;
\item $\it \{tag_{i_j}$ $| i_j \in \{1,\dots,s\}$, $1 \leq j \leq t\}$
is the \emph{set of tags} of the tagged elements $elem_1,\dots,elem_s$;
\item $\it Tag_{i_1},\dots,Tag_{i_t}$ are variables;
\item $\it att_1,\dots,att_n$ are the attribute names;
\item $\it Att_1,\dots,At_n$ are variables, one for each attribute name;
\item $\it NodeTag_{i_1},\dots,NodeTag_{i_t}$ are variables (used for representing the first digit of the node number of the children).
\item $\it NodeTag$ is a variable (used for representing the node number of the tag).
\item $\it k$ is the type number of $\it tag$.
\item $\it r$ is the type number of the tagged elements in $elem_1,\dots,elem_s$ \footnote{Let us remark that given that
{\it tag} is a tagged element then $elem_1,\dots,elem_s$ have been tagged with ``unlabeled" labels when they had
a basic type in the type and node numbered XML document, and thus all of them have a type number.}
\end{itemize}
In addition, we consider \emph{facts} of the form:
\begin{center}
{\it
\begin{tabular}{l}
$\it att_j(v_j,i,k)$ \\
\end{tabular}
}
\end{center}
for each $\it 1 \leq j \leq n$. Finally, for each terminal tagged element in the type and node numbered XML document:
\begin{center}
$\it <tag~nodenumber=i,typenumber=k> value </ tag>$
\end{center}
we consider the \emph{fact}:
\begin{center}
$\it tag(value,i,k).$
\end{center}
In summary, each non-terminal tag (element) is translated into a predicate name, with three arguments.

The first argument of the predicate is used for building a {\it Prolog} term containing the XML document.
It consists of a function symbol
named as $\it ``elementname+type"$ with an argument for each
subelement and an additional argument for storing the list of
attributes.

The second argument of the predicate is used for numbering each node
of the XML document tree, and the third one is use for numbering each type.

Finally, each terminal element and attribute is translated into a
fact. 

Let us remark that the same $\it ``elementname+type"$ function symbol could have several occurrences
with different arity depending on the document includes weakly distinct elements or not.

From a type and node numbered XML document ${\cal X}$,
we can build a unique program ${\cal P}$, and
conversely, from a logic program ${\cal P}$ we can build a unique type and node numbered XML document ${\cal X}$.

The logic program obtained from a document ${\cal X}$ is denoted by $Prog({\cal X})$, and the
XML document obtained from a program ${\cal P}$ is denoted by $Doc({\cal P)}$. In addition, $Doc(Prog({\cal X}))={\cal X}$
and $Prog(Doc({\cal P}))={\cal P}$. 

Moreover, we can associate from our translation to each $tag$
a set of patterns of the form $tagtype(\overline{Tag},[\overline{Att}])$, denoted by $PT(tag)$. 

Finally, to each \emph{pattern} $t$ of $PT(tag)$, we can associate 
 the set of type numbers $\{r_1,\dots,r_n\}$ assigned to $t$ in our translation --there could be more than
 one type number for one pattern due to occurrences of weakly distinct elements--. This set is
denoted by $TN(t)$, and pattern instances 
$t\theta$ have the same set of type numbers, that is,  $TN(t\theta)=_{def}TN(t)$ for all $\theta$.

\subsection{Examples}

For ins\-tance, the running example can be represented by means of
a logic program as follows:
\begin{center}
\noindent {\fontsize{7pt}{9pt}\selectfont
{\it
\begin{tabular}{ll}
\hline
\hspace*{-2.5cm}
\begin{tabular}{l}
Rules (Schema):\\
----------------------------------------\\
books(bookstype(Books, []), NodeBooks,1) :-\\
\hspace*{1cm}book(Books, [NodeBook$|$NodeBooks],2).\\
book(booktype(Author, Title, Review, [Year]), NodeBook ,2) :-        \\
\hspace*{1cm}author(Author, [NodeAuthor$|$NodeBook],3),\\
\hspace*{1cm}title(Title, [NodeTitle$|$NodeBook],3),\\
\hspace*{1cm}review(Review, [NodeReview$|$NodeBook],3),\\
\hspace*{1cm}year(Year, NodeBook,3).\\
review(reviewtype(Unlabeled,Em,[]),NodeReview,3):-\\
\hspace*{1cm}unlabeled(Unlabeled,[NodeUnlabeled$|$NodeReview],4),\\
\hspace*{1cm}em(Em,[NodeEm$|$NodeReview],4).\\
review(reviewtype(Em,[]),NodeReview,3):-\\
\hspace*{1cm}em(Em,[NodeEm$|$NodeReview],5).\\
em(emtype(Unlabeled,Em,[]),NodeEms,5) :-        \\
\hspace*{1cm}unlabeled(Unlabeled,[NodeUnlabeled$|$NodeEms],6),\\
\hspace*{1cm}em(Em, [NodeEm$|$NodeEms],6).\\
\end{tabular} &

\hspace*{-6.75cm}
\begin{tabular}{l}
Facts (Document):\\
------------------------------------------\\
year('2003', [1, 1], 3).\\
author('Abiteboul', [1, 1, 1], 3).\\
author('Buneman', [2, 1, 1], 3).\\
author('Suciu', [3, 1, 1], 3).\\
title('Data on the Web', [4, 1, 1], 3).\\
unlabeled('A', [1, 5, 1, 1], 4).\\
em('fine', [2, 5, 1, 1], 4).\\
unlabeled('book.', [3, 5, 1, 1], 4).\\
year('2002', [2, 1], 3).\\
author('Buneman', [1, 2, 1], 3).\\
title('XML in Scotland', [2, 2, 1], 3).\\
unlabeled('The', [1, 1, 3, 2, 1], 6).\\
em('best', [2, 1, 3, 2, 1], 6).\\
unlabeled('ever!', [3, 1, 3, 2, 1], 6).\\ 
\end{tabular}\\
\hline
\end{tabular}
}
}
\end{center}
Here we can see the translation of each tag
 into a predicate name: $\it books$, $\it book$, etc. Each predicate has three arguments. 
 
The first one, used for representing the
XML document structure, is encapsulated into a
function symbol with the same name as the tag adding the suffix $\it type$. Therefore, we have
$\it bookstype$, $\it booktype$, etc. 

The second argument is used for numbering each node.
For instance, the three facts for the authors of the first book are numbered $\it [1,1,1]$,
$\it [2,1,1]$ and $\it [3, 1, 1]$, representing the authors $\it 'Abiteboul'$, $\it 'Buneman'$ and
$\it 'Suciu'$, respectively, and $\it [1,$ $2, 1]$ for representing $\it 'Buneman'$ in the second book
(see Figure \ref{prole2005}).
Let us remark that the numbering in the facts is in reverse order
with respect to the numbering in the node numbered XML document due to the use of
lists for representing them.

The third argument of the predicate is a number used for numbering each type.
The type number is needed to distinguish weakly distinct elements.
For instance, the tag $\it review$ has two rules, 
one for the case: $\it ``A ~ <\hspace*{-1mm}em\hspace*{-1mm}> ~ fine ~ <\hspace*{-1mm}
/ em\hspace*{-1mm}> ~ book."$
  and other one for the case $\it``<\hspace*{-1mm}em\hspace*{-1mm}> ~The~ <\hspace*{-1mm}em\hspace*{-1mm}> ~best~ 
  <\hspace*{-1mm}/ em\hspace*{-1mm}> ~ ever! ~ <\hspace*{-1mm}/ em\hspace*{-1mm}>"$,
where in the first case the sole emphasized text is $\it 'fine'$, and in
the second case all is emphasized, and
$\it 'best'$ is doubled emphasized. 
The facts and rules in this case are:
\begin{center}
{\fontsize{7pt}{9pt}\selectfont
{\it
\begin{tabular}{l}
\hline
unlabeled('A', [1, 5, 1, 1],  {\bf 4}).\\
em('fine', [2, 5, 1, 1], {\bf 4}).\\
unlabeled('book.', [3, 5, 1, 1], {\bf 4}).\\
unlabeled('The', [1, 1, 3, 2, 1], {\bf 6}).\\
em('best', [2, 1, 3, 2, 1], {\bf 6}).\\
unlabeled('ever!', [3, 1, 3, 2, 1], {\bf 6}).\\
\hline
\end{tabular}
}
}
\end{center}

\begin{center}
{\fontsize{7pt}{9pt}\selectfont
{\it
\begin{tabular}{l}
\hline
review(reviewtype(Unlabeled,Em,[]),NodeReview, {\bf 3}):-\\
\hspace*{1cm}unlabeled(Unlabeled,[NodeUnlabeled$|$NodeReview],{\bf 4}),\\
\hspace*{1cm}em(Em,[NodeEm$|$NodeReview],{\bf 4}).\\
review(reviewtype(Em,[]),NodeReview,{\bf 3}):-\\
\hspace*{1cm}em(Em,[NodeEm$|$NodeReview],{\bf 5}).\\
em(emtype(Unlabeled,Em,[]), NodeEms,{\bf 5}) :-\\
\hspace*{1cm}unlabeled(Unlabeled,[NodeUnlabeled$|$NodeEms],{\bf 6}),\\
\hspace*{1cm}em(Em, [NodeEm$|$NodeEms],{\bf 6}).\\
\hline
\end{tabular}
}
}
\end{center}
They allow us to distinguish
 that the first case is built from the first $\it review$ rule and the second from
the second $\it review$ rule --together with the $\it em$ rule--.
Obviously, in highly non structured documents there could have many schema rules.
The same happens in the case of the following XML document:
\begin{center}
\noindent {\fontsize{7pt}{9pt}\selectfont
{\it
\begin{tabular}{l}
\hline
$<$books$>$\\
$<$book year=$``$2003$"$$>$\\
$<$author$>$Abiteboul$<$/author$>$\\
$<$title$>$Data on the Web$<$/title$>$\\
$<$review$>$A $<$em$>$fine$<$/em$>$ book.$<$/review$>$\\
$<$/book$>$\\
$<$book year=$``$2002$"$$>$\\
$<$author$>$Buneman$<$/author$>$\\
$<$title$>$XML in Scotland$<$/title$>$\\
$<$/book$>$\\
$<$/books$>$\\
\hline
\end{tabular}
}
}
\end{center}
where we have two kinds of records, one with $author$, $title$, $review$ and $year$, and the second one with
$author$, $title$ and $year$. In this case, we have to consider the following schema rules:
\begin{center}
{\fontsize{7pt}{9pt}\selectfont
{\it
\begin{tabular}{l}
\hline
books(bookstype(Book, []), NodeBooks,{\bf 1}):-\\
\hspace*{1cm}book(Book, [NodeBook$|$NodeBooks],{\bf 2}).\\
book(booktype(Author, Title, Review, [Year]), NodeBook,{\bf 2}) :-\\
\hspace*{1cm}author(Author, [NodeAuthor$|$NodeBook],{\bf 3}),\\
\hspace*{1cm}title(Title, [NodeTitle$|$NodeBook],{\bf 3}),\\
\hspace*{1cm}review(Review, [NodeReview$|$NodeBook],{\bf 3}),\\
\hspace*{1cm}year(Year, NodeBook,{\bf 3}).\\
book(booktype(Author, Title, [Year]), NodeBook,{\bf 2}) :- \\
\hspace*{1cm}author(Author, [NodeAuthor$|$NodeBook],{\bf 4}),\\
\hspace*{1cm}title(Title, [NodeTitle$|$NodeBook],{\bf 4}),\\
\hspace*{1cm}year(Year, NodeBook,{\bf 4}).\\
author('Abiteboul',[1,1,1],{\bf 3}).\\
author('Buneman',[1,2,1],{\bf 4}).\\
...\\
\hline
\end{tabular}
}
}
\end{center}
The use of numbers {\bf 2-3-3-3-3} and {\bf 2-4-4-4} in the above rules,
and in the corresponding facts, allows the distinction of the subelements
of $\it Abiteboul$ and $\it Buneman$'s books. The use of the same type numbering would suppose 
ambiguity, given that the $\it Abiteboul$'s book has also the type described by second rule of $\it book$.

On the other hand, whenever in a tagged element there is more than one value for the same
subtag, we introduce one fact for each value, numbered with the same type number, but distinct node number.
For instance, with respect to the running example:
\begin{center}
{\fontsize{7pt}{9pt}\selectfont
{\it
\begin{tabular}{l}
\hline
author('Abiteboul', {\bf [1, 1, 1], 3}).\\
author('Buneman', {\bf [2, 1, 1], 3}).\\
author('Suciu', {\bf [3, 1, 1], 3}).\\
\hline
\end{tabular}
}
}
\end{center}

In addition, the attributes of tagged elements are stored in a {\it Prolog} list.
For instance, with respect to the following XML document:
\begin{center}
{\fontsize{7pt}{9pt}\selectfont
{\it
\begin{tabular}{l}
\hline
$<$book year=$``$2003$"$,keyword=$``$XML$"$$>$\\
$<$author$>$Abiteboul$<$/author$>$\\
$<$title$>$Data on the Web$<$/title$>$\\
$<$review$>$A $<$em$>$fine$<$/em$>$ book.$<$/review$>$\\
$<$/book$>$\\
\hline
\end{tabular}
}
}
\end{center}
we will consider the following schema rule:
\begin{center}
{\fontsize{7pt}{9pt}\selectfont
{\it
\begin{tabular}{l}
\hline
book(booktype(Author, Title, Review, {\bf [Year,Keyword]}), NodeBook, 2) :-\\
\hspace*{1cm}author(Author, [NodeAuthor$|$NodeBook],3),\\
\hspace*{1cm}title(Title, [NodeTitle$|$NodeBook],3),\\
\hspace*{1cm}review(Review, [NodeReview$|$NodeBook],3),\\
\hspace*{1cm}year({\bf Year},NodeBook,3),\\
\hspace*{1cm}keyword({\bf Keyword},NodeBook,3).\\
\hline
\end{tabular}
}
}
\end{center}

Finally, each value in a non-terminal tagged element
is translated into a fact called $\it unlabeled$. This is the
case in the running example of $\it 'A'$ and $\it 'book.'$ in the first review, 
and $\it 'The'$ and $\it 'ever!'$ in the second one.

\vspace*{-3mm}
\section{Program Specialization for XPath Expressions}

In this section, we will present the program specialization
technique for querying {\it XPath} expressions against an XML document
represented by means of a logic program. Firstly, we present the semantic of the
{\it XPath} expressions.

\subsection{XPath Semantics}

An \emph{XPath expression} $xpathexpr$ has the form $/expr_1/\dots /expr_n$ where each \emph{simple
XPath expression} $expr_i$ has the form:

\begin{enumerate}
\item $expr \equiv tag$
\item $expr \equiv tag[cond]$
\item $expr \equiv @att$
\item $expr \equiv text()$
\end{enumerate}

and $cond$ is a boolean condition which has the form:

\begin{enumerate}
\item[(a)] $cond \equiv tag=value$
\item[(b)] $cond \equiv @att=value$
\item[(c)] $cond \equiv cond_1~and~cond_2$
\item[(d)] $cond \equiv cond_1~or~cond_2$
\item[(e)] $cond \equiv xpathexpr$ 
\end{enumerate}

The above expressions $expr_i$ when $1 \leq i < n$ can only be chosen from the
cases (1) and (2). We consider only a subset of {\it XPath} w.r.t. the
{\it XPath} specification \cite{XPath} which can specify paths on XML trees 
and restricts boolean conditions to express equalities to values 
connected with {\it ``and"} and {\it ``or"} logic connectives. This restriction is
enough to understand our proposed technique. More complex
{\it XPath} queries can be translated into logic programming following 
similar ideas. We have implemented in our prototype a rich set
of {\it XPath} queries including primitives ``*", ``//", ``/../" , ``$>$",``$<$", etc. 

The semantics of the previous {\it XPath} expressions is as follows. Given
an XML document, an {\it XPath} expression defines a subtree of the XML
document. It can be defined as the subtree obtained from the XML tree satisfying each simple
expression $expr$ in the {\it XPath} expression.The semantics of {\it XPath} expressions could
be defined as a forest (i.e. a sequence of subtrees) instead of a tree. However, we have adopted this
definition in which an {\it XPath} expression defines a rooted document. 
The root is the same as the input document and therefore describes a {\it complete branch of
the input document}. More
concretely:

Given an XML document ${\cal X}$ and an {\it XPath} expression $xpathexpr=/expr_r$ $\dots $ $/expr_n$
the \emph{subtree  of ${\cal X}$ defined by $xpathexpr$} is denoted by $subtree({\cal X},xpathexpr)$
and defined as:

\begin{itemize}
\item[(a)] If ${\cal X}$ is a non terminal tagged element and has the form
$$<tag~att_1=v_1,\dots,att_n=v_n > elem_1,\dots,elem_s </ tag>$$
then\\
(a.1):
$$\begin{array}{l}
subtree({\cal X},/expr_r/ \dots /expr_n)=_{def}\\
~~~~~~~~~<tag~att_1=v_1,\dots,att_n=v_n >\\
~~~~~~~~~subtree(elem_{1},/expr_{r+1}/ \dots /expr_n),\\
~~~~~~~~~\dots,\\
~~~~~~~~~subtree(elem_{s},/expr_{r+1}/ \dots /expr_n),\\
~~~~~~~~~elem_{i_1},\\
~~~~~~~~~\dots,\\
~~~~~~~~~elem_{i_k}\\
~~~~~~~~~</ tag>~\\
\end{array}$$
whenever $r<n$ and ${\cal X}$ satisfies $expr_r$; 
where $elem_{i_1},\dots,elem_{i_k}$  is the
subsequence of $elem_1,\dots,elem_s$
satisfying $cond$ whenever $expr_r \equiv tag[cond]$;
(a.2):
$$\begin{array}{l}
subtree({\cal X},/expr_n)=_{def} {\cal X}\\
\end{array}$$
whenever $r=n$ and ${\cal X}$ satisfies $expr_n$; and\\
(a.3):
$$subtree({\cal X},/expr_r/ \dots /expr_n)=_{def} \epsilon$$ otherwise.
\item[(b)] If ${\cal X}$ is a terminal tagged element then\\
(b.1):
$$subtree({\cal X},/expr_r/ \dots /expr_n)=_{def} {\cal X}$$
whenever $r=n$ and ${\cal X}$ satisfies $expr_r$; and\\
(b.2):
$$subtree({\cal X},/expr_r/ \dots /expr_n)=_{def} \epsilon$$ otherwise.
\item[(c)] If ${\cal X}$ has a basic type then\\
(c.1): $$subtree({\cal X}, /text())=_{def}{\cal X}$$ and\\
(c.2): $$subtree({\cal X}, xpathexpr)=_{def}\epsilon$$
whenever $xpathexpr \not\equiv /text()$
\end{itemize}
where $\epsilon$ denotes the empty sequence.

In addition, an XML document  ${\cal X}$ \emph{satisfies a simple XPath expression} $expr$
in the following cases:\\

\noindent (i) $ \hspace*{5mm}{\cal X} \equiv <tag~att_1=v_1,\dots,att_n=v_n > elem_1,\dots,elem_s </ tag>$\\

\noindent satisfies $expr$ whenever:

\begin{enumerate}
\item[(i.1)] $expr \equiv tag$
\item[(i.2)] $expr \equiv tag[cond]$
and ${\cal X}$ satisfies the condition $cond$, that is:
\begin{enumerate}
\item[(i.2.1)] $cond \equiv tag'=value$ and $tag'$ is a terminal tagged subelement of $tag$ and the value
of $tag'$ is equal to $value$.
\item[(i.2.2)] $cond \equiv @att=value$, some $att_i$ $1 \leq i \leq n$ is equal to $att$, 
and $v_i$ is equal to $value$.
\item[(i.2.3)] $cond \equiv cond_1~and~cond_2$, ${\cal X}$ satisfies the condition $cond_1$ and 
${\cal X}$ satisfies the condition $cond_2$.
\item[(i.2.4)] $cond \equiv cond_1~or~cond_2$, ${\cal X}$ satisfies the condition $cond_1$ or 
${\cal X}$ satisfies the condition $cond_2$.
\item[(i.2.5)] $cond \equiv xpathexpr$ and $subtree({\cal X},/tag/xpathexpr)$ is a branch of ${\cal X}$.
\end{enumerate}
\item[(i.3)] $expr \equiv @att$ and some $att_i$ $1 \leq i \leq n$ is equal to $att$
\end{enumerate}
and\\

\noindent (ii) $ \hspace*{5mm} {\cal X}~has~a~basic~type$\\

\noindent satisfies $expr$ whenever $expr \equiv text()$.\\

\noindent For instance, w.r.t. the running example, 
the {\it XPath} expression $\it /books/book[author$ $= ``Suciu"]/title$ defines 
$subtree({\cal X},/books/book[author = ``Suciu"]/title)$ which is equal to:

\begin{center}
{\fontsize{7pt}{9pt}\selectfont
{\it
\noindent \begin{tabular}{l}
\hline
$<$books $>$\\
subtree(${\cal X'}$,/book[author=``Suciu"]/title)\\
subtree(${\cal X''}$,/book[author=``Suciu"]/title)\\
$<$/books$>$\\
\hline
\end{tabular}
}
}
\end{center}

\noindent by case (a.1) of the definition, since there is no boolean conditions
in $books$, where ${\cal X'}$ is:

\begin{center}
{\fontsize{7pt}{9pt}\selectfont
{\it
\noindent \begin{tabular}{l}
\hline
$<$book year=$``$2003$"$$>$\\
$<$author$>$Abiteboul$<$/author$>$\\
$<$author$>$Buneman$<$/author$>$\\
$<$author$>$Suciu$<$/author$>$\\
$<$title$>$Data on the Web$<$/title$>$\\
$<$review$>$A $<$em$>$fine$<$/em$>$ book.$<$/review$>$\\
$<$/book$>$\\
\hline
\end{tabular}
}
}
\end{center}

\noindent and ${\cal X''}$ is:

\begin{center}
{\fontsize{7pt}{9pt}\selectfont
{\it
\noindent \begin{tabular}{l}
\hline

$<$book year=$``$2002$"$$>$\\
$<$author$>$Buneman$<$/author$>$\\
$<$title$>$XML in Scotland$<$/title$>$\\
$<$review$>$$<$em$>$The $<$em$>$best$<$/em$>$ ever!$<$/em$>$$<$/review$>$\\
$<$/book$>$\\
\hline
\end{tabular}
}
}
\end{center}

\noindent In addition, $subtree({\cal X'},/book[author=``Suciu"]/title)$ is equal to:

\begin{center}
{\fontsize{7pt}{9pt}\selectfont
{\it
\noindent \begin{tabular}{l}
\hline
$<$book year=``2003"$>$\\
$<$author$>$Suciu$<$/author$>$\\
subtree(${\cal X'''}$,/title)\\
$<$/book$>$\\
\hline
\end{tabular}
}
}
\end{center}

\noindent by case (a.1) of the definition, given that the boolean condition $[author=``Suciu"]$ is satisfied by $<author>Suciu</author>$,
by case (i.2.1) of definition, and is not satisfied by $<author>Abiteboul</author>$ and $<author>Buneman</author>$.
In addition, ${\cal X}'''$ is:

\begin{center}
{\fontsize{7pt}{9pt}\selectfont
{\it
\noindent \begin{tabular}{l}
\hline
$<$title$>$XML in Scotland$<$/title$>$\\
\hline
\end{tabular}
}
}
\end{center}

\noindent and $subtree({\cal X''},/book[author=``Suciu"]/title)=\epsilon$, by case (a.3) of the definition. Finally,
$subtree({\cal X'''},/title)$ is equal to:

\begin{center}
{\fontsize{7pt}{9pt}\selectfont
{\it
\noindent \begin{tabular}{l}
\hline
$<$title$>$XML in Scotland$<$/title$>$\\
\hline
\end{tabular}
}
}
\end{center}

\noindent by case (a.2) of the definition. Therefore $subtree({\cal X},/books/book[author=``Suciu"]/$ $title)$ is equal to:

\begin{center}
{\fontsize{7pt}{9pt}\selectfont
{\it
\noindent \begin{tabular}{l}
\hline
$<$books$>$\\
$<$book year=$``$2003$"$$>$\\
$<$author$>$Suciu$<$/author$>$\\
$<$title$>$XML in Scotland$<$/title$>$\\
$<$/book$>$\\
$<$/books$>$\\
\hline
\end{tabular}
}
}
\end{center}

In other words, the subtree defined by an {\it XPath} expression can be seen as the subtree of the input XML document
which is traversed for answering the query. In practice, the answer to an {\it XPath} query consists of the 
sequence of subtrees (i.e. the forest) of the tree defined by the {\it XPath} expression, 
whose tag is equal to the rightmost tag of the {\it XPath} query. For instance,
in the above example, the answer would be:

\begin{center}
{\fontsize{7pt}{9pt}\selectfont
{\it
\noindent \begin{tabular}{l}
\hline
$<$title$>$XML in Scotland$<$/title$>$\\
\hline
\end{tabular}
}
}
\end{center}

\noindent given that the rightmost tag of the {\it XPath} query is $title$.

\subsection{Schema Rule Specialization}

The \emph{first step of the program specialization} consists of a predicate removing from the schema rules.

With this aim, we need to map each {\it XPath} expression to a so-called \emph{free of equalities XPath expression}.
Each {\it XPath} expression $xpathexpr=/expr_1 \dots /expr_n$ can be mapped into a
\emph{free of equalities XPath expression} as follows. 

Each simple {\it XPath} expression
$expr$ can be mapped into a \emph{free of equalities simple XPath expression}
denoted by $FE(expr)$.  Analogously, we need to define $FE(cond)$ which is
a \emph{free of equalities boolean condition} associated to a boolean condition $cond$.
They are defined as follows, distinguishing cases in the form of $expr$ and
$cond$.

\begin{enumerate}
\item $expr \equiv tag$: $FE(expr)=_{def} expr.$
\item $expr \equiv tag[cond]$: $FE(expr)=_{def}tag[FE(cond)]$
\item $expr \equiv @att$: $FE(expr)=_{def}@att$
\item $expr \equiv text()$: $FE(expr)=_{def}text()$
\item $cond \equiv tag=value$: $FE(expr)=_{def}tag$
\item $cond \equiv @att=value$: $FE(expr)=_{def}@att$
\item $cond \equiv cond_1~and~cond_2$: $FE(expr)=_{def}FE(cond_1)~ and~ FE(cond_2)$
\item $cond \equiv cond_1~or~cond_2$: $FE(expr)=_{def}FE(cond_1)~ or~ FE(cond_2)$
\item $cond \equiv xpathexpr$: $FE(expr)=_{def}FE(xpathexpr)$
\end{enumerate}

Now, given $xpathexpr=/expr_1/ \dots /expr_n$ then
$FE(xpathexpr)=_{def}/FE($ $expr_1)/ $ $\dots /FE(expr_n)$.  
Free of equalities {\it XPath} expressions $xpathfree$ are expressions $/fexpr_1/
\dots /fexpr_n$ where each $fexpr_i$, $1 \leq i \leq n$, has the form:

\begin{enumerate}
\item $fexpr \equiv tag$
\item $fexpr \equiv tag[cond]$
\item $fexpr \equiv @att$
\item $fexpr \equiv text()$
\end{enumerate}
and $cond$ is a free of equalities boolean condition which has the form:
\begin{enumerate}
\item[(a)] $cond \equiv cond_1~and~cond_2$
\item[(b)] $cond \equiv cond_1~or~cond_2$
\item[(c)] $cond \equiv xpathfree$
\end{enumerate}

Free of equalities {\it XPath} expressions define a subtree of the
XML document in which some subpaths of the XML document must exist
due to occurrences of free of equalities boolean conditions.  

For instance, in the running example, $\it FE(/books/book$ $\it [author = ``Suciu" ]/title)$ $=$
$\it /books/book$ $\it [author]$ $/title$,
and the subtree of the (type and node numbered) XML document which corresponds with the {\it XPath}
expression $\it /books/book$ $\it [author]$ $/title$ is as follows:
\begin{center}
{\fontsize{7pt}{9pt}\selectfont
{\it
\noindent \begin{tabular}{l}
\hline
$<$books nodenumber=1, typenumber=1$>$\\
$<$book year=$``$2003$"$, nodenumber=1.1, typenumber=2$>$\\
$<$author nodenumber=1.1.1 typenumber=3$>$Abiteboul$<$/author$>$\\
$<$author nodenumber=1.1.2 typenumber=3$>$Buneman$<$/author$>$\\
$<$author nodenumber=1.1.3 typenumber=3$>$Suciu$<$/author$>$\\
$<$title nodenumber=1.1.4 typenumber=3$>$Data on the Web$<$/title$>$\\
$<$/book$>$\\
$<$book year=$``$2002$"$ nodenumber=1.2, typenumber=2$>$\\
$<$author nodenumber=1.2.1 typenumber=3$>$Buneman$<$/author$>$\\
$<$title nodenumber=1.2.2 typenumber=3$>$XML in Scotland$<$/title$>$\\
$<$/book$>$\\
$<$/books$>$\\
\hline
\end{tabular}
}
}
\end{center}
Let us remark that the boolean condition $\it [author]$
forces to include each author in the subtree
represented by the free of equalities {\it XPath} expression $\it /books/book$ $[author]/$ $\it title$.

Now, given a type and node numbered XML document ${\cal X}$ and an {\it XPath} expression $xpathexpr$ then
the \emph{specialized program ${\cal P}^{xpathexpr}$ obtained from ${\cal P}$} is defined as 
the schema rules for the subtree of ${\cal X}$ defined by $xpathfree$, where $xpathfree$ is the free of equalities
{\it XPath} expression obtained from $xpathexpr$, together with the facts of ${\cal P}$. In other words: 
$${\cal P}^{xpathexpr}=_{def} Rules(Prog(subtree({\cal X},FE(xpathexpr)))) \cup Facts({\cal P})$$

For instance, with respect to the running example and $/books/book$ $\it [author = ``Suciu" ]/title$,
${\cal P}^{/books/book[author = ``Suciu" ]/title}$ consists of the specialized schema rules:

\begin{center}
{\fontsize{7pt}{9pt}\selectfont
{\it
\noindent \begin{tabular}{l}
\hline
books(bookstype(Books, []), NodeBooks,1):-\\
\hspace*{1cm}book(Book, [NodeBook$|$NodeBooks],2).\\
book(booktype(Author,Title,Review,[Year]),NodeBook,2) :-\\
\hspace*{1cm}author(Author,[NodeAuthor$|$NodeBook],3),\\
\hspace*{1cm}title(Title,[NodeTitle$|$NodeBook],3). \\
\hline
\end{tabular}
}
}
\end{center}

\noindent together with the set of facts of ${\cal P}$.

Let us remark that in practice, the specialized schema rules can be
obtained from the schema rules by removing predicates; that is,
removing the predicates in the schema rules which are not tags in
the (free of equalities) {\it XPath} expression.  

\subsection{Generation of Goals}

The \emph{second step of the specialization program} consists of 
(1) to consider the equalities removed from the original {\it XPath} expression
when the free of equalities {\it XPath} expression was generated, and (2) to
generate a set of goals from these equalities.
  
With this aim, each {\it XPath} expression $xpathexpr$ can be mapped into a set of {\it Prolog} terms, denoted by $PT(xpathexpr)$,
denoting the set of \emph{patterns of the query}. These patterns are instances of the {\it ``elementname+type"}
patterns defined in our translation. 

In particular, each simple {\it XPath} expression $expr$ can be mapped into a set of patterns, denoted
by $PT(expr)$.  This set can be defined as follows, distinguishing cases in the form of $expr$:

\begin{enumerate}
\item $expr \equiv tag$: $PT(expr)=_{def}\emptyset.$
\item $expr \equiv tag[cond]$:
\begin{enumerate}
\item $cond \equiv tag_i=value$:
$PT(expr)=_{def}\{tagtype(\overline{Tag},[\overline{Att}]) \{ Tag_i \rightarrow value\} |$
$tagtype(\overline{Tag},[\overline{Att}]) \in PT(tag)\}$.
\item $cond \equiv @att_i=value$:
$PT(expr)=_{def}\{ tagtype(\overline{Tag},[\overline{Att}])
\{ Att_i \rightarrow value\} |$
$tagtype(\overline{Tag},[\overline{Att}]) \in PT(tag)\}$.
\item $cond \equiv cond_1~and~cond_2$. $PT(expr)=_{def}\{t\theta | \theta=m.g.u.(t,t'), t \in PT(tag$ $[cond_1]),$ $t' \in PT(tag[cond_2]))\}$
\item $cond \equiv cond_1~or~cond_2$. $PT(expr)=_{def}PT(tag$ $[cond_1]) \cup PT(tag[cond_2])$
\item $cond \equiv xpathexpr$: $PT(expr)=_{def}PT(xpathexpr)$
\end{enumerate}
\item $expr \equiv @att$: $PT(expr)=_{def}\emptyset$
\item $expr \equiv text()$: $PT(expr)=_{def}\emptyset$
\end{enumerate}

Now, $$PT(/expr_1/ \dots /expr_n)=_{def} \{t_1\theta | \theta=m.g.u.(t_1,\dots,t_n), t_i \in PT(expr_i), 1 \leq i \leq n\}$$

Now, given a type and node numbered XML document and an {\it XPath} expression $xpathexpr$ then 
the \emph{set of specialized goals
for $xpathexpr$} is defined as the set:
$${\cal G}^{xpathexpr}=_{def}$$ 
$$\{tag(Pattern,Node,Type)\{Pattern \rightarrow t,Type \rightarrow r\}~|$$ $$~t \in PT(xpathexpr),r \in TN(t)\}$$
where $tag$ is the leftmost tag in $xpathexpr$ with a boolean condition.
If there is no boolean conditions, the set is defined as: 
$${\cal G}^{xpathexpr}=_{def}$$
$$\{tag(Pattern,Node,Type)\{Type \rightarrow r\} |$$ $$t \in PT(tag),r \in TN(t)\}$$

For instance, with respect to $\it /books/book$ $\it [author = ``Suciu" ]/title$
and the running example $\it PT(/books/$ $\it book[author = ``Suciu" ]/title)=\{booktype('Suciu', Title,$ $Review, $ $[Year])\}$
and $TN(booktype('Suciu', Title, Review,$ $[Year]))=\{2\}$. Therefore the (unique) goal is $\it :-
book(booktype('Suciu', Title, Review, [Year]), Node, 2)$.

In summary, the handling of an {\it XPath} query involves the specialization of the schema rules of the XML document and
the generation of one or more goals. The goals are obtained from the leftmost tag in the {\it XPath} expression
with a boolean condition, instantiated by mean of patterns obtained from the boolean equalities.

 \vspace*{-2mm}
\subsection{Reconstruction of the answer}
\vspace*{-1mm}
In order to rebuild the answer, we have to reason as follows. 

A logic program ${\cal P}$ obtained from
an XML document ${\cal X}$ contains schema rules and facts of the form $att(value,i,r)$ and $tag(value,i,r)$,
and conversely, from this set of facts and the schema rules we can rebuild the document ${\cal X}$.

However, the same (and fragments of the) XML document ${\cal X}$ can be also obtained from the schema rules and facts of the form $att(value,i,r)$ and $tag(t,i,r)$ whenever $t$'s are {\it Prolog} terms of the form $tagtype(s,j,k)$,
--$t$ are pattern instances-- and $tag(t,i,r)$ belongs to the Herbrand model (with variables) of ${\cal P}$.

\noindent For instance, from the following fact:

\vspace*{-1mm}
\begin{center}
{\fontsize{7pt}{9pt}\selectfont
{\it
\noindent \begin{tabular}{l}
\hline
book(booktype('Abiteboul', Title, reviewtype('A    ', fine, []),['2003']),[1,1],2).\\
\hline
\end{tabular}
}
}
\end{center}
and the schema rules of the running example, we can rebuild the XML document:
\begin{center}
{\fontsize{7pt}{9pt}\selectfont
{\it
\noindent \begin{tabular}{l}
\hline
$<$books nodenumber=1, typenumber=1$>$\\
$<$book year=$``$2003$"$, nodenumber=1.1, typenumber=2$>$\\
$<$author nodenumber=1.1.1 typenumber=3$>$Abiteboul$<$/author$>$\\
$<$review nodenumber=1.1.5 typenumber=3$>$\\
$<$unlabeled nodenumber=1.1.5.1 typenumber=4$>$ A $<$/ unlabeled$>$\\
$<$em nodenumber=1.1.5.2 typenumber=4$>$fine$<$/em$>$\\
$<$/review$>$\\
$<$/book$>$\\
$<$/books$>$\\
\hline
\end{tabular}
}
}
\end{center}
\vspace*{-2mm}
Let us remark that the previous fact represents a fragment of the whole XML document, where 
the type and node numbering together with the schema rules allow us to rebuild this fragment of the XML document. In this fact the variable $Title$ represents a missing value in the XML document.

Therefore when a goal obtained from an {\it XPath} expression is called, 
each answer of the goal represents a fragment of the {\it XPath} query answer.

Given a type and node numbered XML document ${\cal X}$, 
the logic program ${\cal P}$ representing ${\cal X}$,
and an {\it XPath} expression $xpathexpr$,
then we can build the \emph{XML document representing the answer}, denoted by $Doc(xpathexpr,{\cal P})$, as follows:
\begin{center}
\begin{tabular}{l}
$Doc(xpathexpr,{\cal P})=_{def} Doc(Rules({\cal P}^{xpathexpr}) \cup$\\
\hspace*{0.35cm} $  \{tag(t,Node,r)\theta | \theta$ is an answer of $tag(t,Node,r)$,\\
\hspace*{0.5cm} w.r.t. ${\cal P}^{xpathexpr}$, $tag(t,$ $Node,r) \in {\cal G}^{xpathexpr}$ $\})$\\
\end{tabular}
\end{center}

\noindent Analogously, when the {\it XPath} expression $xpathexpr$ has no boolean conditions:

\begin{center}
\begin{tabular}{l}
$Doc(xpathexpr,{\cal P})=_{def} Doc(Rules({\cal P}^{xpathexpr}) \cup$\\
\hspace*{0.35cm} $  \{tag(X,Node,r)\theta | \theta$ is an answer of $tag(X,Node,r)$,\\
\hspace*{0.5cm} w.r.t. ${\cal P}^{xpathexpr}$, $tag(X,Node,r) \in {\cal G}^{xpathexpr}$ $\})$\\
\end{tabular}
\end{center}

Let remark us that our programs have finite answers and thus the previous definition
has sense. In addition, the previous definition defines the XML document answer of an {\it XPath} expression
as a \emph{complete branch of the input XML document}. 

For instance, w.r.t. the running example and the {\it XPath} expression $/books/book$ $[author$ $=``Suciu"]/title$,
the (unique) goal is $:-book(booktype('Suciu',Title,$ $Review,$ $[Year],Node,2)$, and the (unique)
answer of the goal w.r.t. the following specialized schema rule:

\begin{center}
{\fontsize{7pt}{9pt}\selectfont
{\it
\noindent \begin{tabular}{l}
\hline
book(booktype(Author,Title,Review,[Year]),NodeBook,2) :-\\
\hspace*{1cm}author(Author,[NodeAuthor$|$NodeBook],3),\\
\hspace*{1cm}title(Title,[NodeTitle$|$NodeBook],3). \\
\hline
\end{tabular}
}
}
\end{center}

\noindent is {\it $\theta=\{$Title $/$ $'$Data on the Web$'$, Node $/$ [1,1] $\}$}. Now, from the goal instance
{\it book(booktype($'$Suciu$'$, $'$Data on the Web$'$, Review, [Year], [1,1], 2)} obtained from $\theta$,
we can rebuild the answer:

\begin{center}
{\fontsize{7pt}{9pt}\selectfont
{\it
\noindent \begin{tabular}{l}
\hline
$<$books nodenumber=1, typenumber=1$>$\\
$<$book  nodenumber=1.1, typenumber=2$>$\\
$<$author nodenumber=1.1.1 typenumber=3$>$Suciu$<$/author$>$\\
$<$title nodenumber=1.1.4 typenumber=3$>$Data on the Web$<$/title$>$\\
$<$/book$>$\\
$<$/books$>$\\
\hline
\end{tabular}
}
}
\end{center}

Therefore, the XML document representing the answer of an {\it XPath} expression is defined as
the document obtained from the specialized schema rules and the
goal instances obtained from each answer of the goals.  

\subsection{Reordering}

Finally, there is an optimization in our proposed technique which consists in 
the reordering of predicates in the schema rules in order
to follow a \emph{left-to-right evaluation order} of {\it XPath} expressions. 
The aim of such left-to-right evaluation order is to keep the order of filtering 
that the user specifies by means of the boolean conditions. 

For instance, 
in the case  of the {\it XPath} expression $\it /books/book[@year=2002 ~$ $\it and~title$ $\it =``Data~on~the~Web"]/author$,
the user has required the authors of the books published in the year $\it 2002$ with title {\it ``Data on the Web"}.
Following a left-to-right evaluation order, firstly, the books are filtered by the year, and after by the title.

This predicate reordering is as follows.
Supposing the {\it XPath} expression $\it /books$ $/book$ $[@year=\it 2002 ~$ $\it and~title$ $\it =``Data~on~the~Web"]/author$,
the schema rule specialization should correspond with:
\begin{center}
{\fontsize{7pt}{9pt}\selectfont
{\it
\begin{tabular}{l}
\hline
book(booktype(Author, Title, Review, [Year]),NodeBook,2):-\\
\hspace*{1cm}author(Author, [NodeAuthor$|$NodeBook],3),\\
\hspace*{1cm}title(Title,[NodeTitle$|$NodeBook],3),\\
\hspace*{1cm}year(Year,NodeBook,3).\\
\hline
\end{tabular}
}
}
\end{center}
However, in order to follow a left-to-right evaluation order of the {\it XPath} expression, we reorder the predicates in the body
of the predicate $book$ and we transform this schema rule into:
\begin{center}
{\fontsize{7pt}{9pt}\selectfont
{\it
\begin{tabular}{l}
\hline
book(booktype(Author, Title, Review, [Year]),NodeBook,2):-\\
\hspace*{1cm}year(Year,NodeBook,3),\\
\hspace*{1cm}title(Title,[NodeTitle$|$NodeBook],3),\\
\hspace*{1cm}author(Author, [NodeAuthor$|$NodeBook],3).\\
\hline
\end{tabular}
}
}
\end{center}
in which, firstly, the books are filtered by year, after the titles are obtained, and finally, the authors are computed.

\subsection{Examples}

In this section we would like to show some examples of the proposed technique.
In each example, we will show the specialized schema rules, the set of generated goals, the set of answers, and the answer in the form of an XML document obtained from the goal instances.

\subsubsection*{Example 1} 
For instance, we can suppose an {\it XPath} query such as
$\it /books/book/author$, requiring the authors in the book database. In this case, we have to consider the unique goal
$\it :-author(Author,Node,3)$, given that $PT(author)=\{authortype(Author,[])\}$ and
$TN(authortype(Author,[]))=\{3\}$. The call of such a goal will compute the answers:\\
{\fontsize{7pt}{9pt}\selectfont
{\it
\begin{tabular}{ll}
\hline
(1) {\it Author/'Abiteboul'} & {\it Node/[1,1,1]}\\
(2) {\it Author/'Buneman'} & {\it Node/[2,1,1]}\\
(3) {\it Author/'Suciu'} & {\it Node/[3,1,1]}\\
(4) {\it Author/'Buneman'} & {\it Node/[1,2,1]}\\
\hline
\end{tabular}
}
}
\noindent which correspond with the following set of goal instances and XML document:\\
{\fontsize{7pt}{9pt}\selectfont
{\it
 \begin{tabular}{cc}
\hline
\hspace*{-2cm}\begin{tabular}{l}
author('Abiteboul', [1, 1, 1],3).\\
author('Buneman', [2, 1, 1],3).\\
author('Suciu', [3, 1, 1],3).\\
author('Buneman', [1, 2, 1],3).\\
\end{tabular} &
\hspace*{-9cm}\begin{tabular}{l}
$<$result$>$\\
$<$author$>$Abiteboul$<$/author$>$\\
$<$author$>$Buneman$<$/author$>$\\
$<$author$>$Suciu$<$/author$>$\\
$<$author$>$Buneman$<$/author$>$\\
$<$/result$>$\\
\end{tabular}\\
\hline
\end{tabular}
}
}
Let us remark that answer is packed into a tag called $result$.

\subsubsection*{Example 2} 
Now, we can suppose the {\it XPath} expression $\it /books/book$. 
Now, the unique goal is $\it :-book(Book,Node,2)$, because $PT(book)=\{booktype(Author,Title,Review,[Year$ $])\}
$ and $TN(booktype(Author,Title,Review,[Year]))=\{2\}$.
 The call of the goal $\it book($ $Book,$ $Node,2)$ computes the following answers:\\
{\fontsize{7pt}{9pt}\selectfont
{\it
 \begin{tabular}{c}
\hline
(1) {\it Book/booktype('Abiteboul', 'Data on the Web', reviewtype('A', 'fine', []),['2003'])}\\ {\it Node/[1, 1]}\\
(2) {\it Book/booktype('Abiteboul', 'Data on the Web', reviewtype('book.', 'fine', []), ['2003'])}\\ {\it Node/[1, 1]}\\
(3) {\it Book/booktype('Buneman', 'Data on the Web', reviewtype('A', 'fine', []), ['2003'])}\\ {\it Node/[1, 1]}\\
(4) {\it Book/booktype('Buneman', 'Data on the Web', reviewtype('book.', 'fine', []), ['2003'])}\\ {\it Node/[1, 1]}\\
(5) {\it Book/booktype('Suciu', 'Data on the Web', reviewtype('A', 'fine', []), ['2003'])}\\ {\it Node/[1, 1]}\\
(6) {\it Book/booktype('Suciu', 'Data on the Web', reviewtype('book.', 'fine', []), ['2003'])}\\ {\it Node/[1, 1]}\\
(7) {\it Book/booktype('Buneman', 'XML in Scotland', reviewtype(emtype('The', 'best', []), []), ['2002'])}\\ {\it Node/[2, 1]}\\
(8) {\it Book/booktype('Buneman', 'XML in Scotland', reviewtype(emtype('ever!', 'best', []), []), ['2002'])}\\ {\it Node/[2, 1]}\\
\hline
\end{tabular}
}
}
\noindent which corresponds with the following document:\\
{\fontsize{7pt}{9pt}\selectfont
{\it
 \begin{tabular}{l}
\hline
$<$result$>$\\
$<$book year=$``$2003$"$$>$\\
$<$author$>$Abiteboul$<$/author$>$\\
$<$author$>$Buneman$<$/author$>$\\
$<$author$>$Suciu$<$/author$>$\\
$<$title$>$Data on the Web$<$/title$>$ \\
$<$review$>$\\
A $<$em$>$fine$<$/em$>$ book.\\
$<$/review$>$\\
$<$/book$>$\\
$<$book year=$``$2002$"$$>$\\
$<$author$>$Buneman$<$/author$>$\\
$<$title$>$XML in Scotland$<$/title$>$\\
$<$review$>$\\
$<$em$>$ The $<$em$>$best$<$/em$>$ ever!$<$/em$>$\\
$<$/review$>$\\
$<$/book$>$\\
$<$/result$>$\\
\hline
\end{tabular}
}
}
 
 \subsubsection*{Example 3} 
 Let us consider the {\it XPath} expression
 $\it /books/book$ $\it [author = ``Suciu" ]/title$. In this case, we have a condition in
 the form of $\it author=``Suciu"$. 

Therefore we have to consider (a) the goal $\it
:-book(booktype('Suciu', Title, Review,$ $[Year]), Node, 2)$
given that $PT(/books/book$ $\it [author = ``Suciu" ]/title)=\{booktype$ $('Suciu', Title, Review, [Year])\}$ and
$TN(booktype('Suciu', Title, Review, [Year]))$ $=\{2\}$; and we have
to consider (b) the following specialized rule:\\
{\fontsize{7pt}{9pt}\selectfont
{\it
\begin{tabular}{l}
\hline
book(booktype(Author,Title,Review,[Year]),NodeBook,2) :-\\
\hspace*{1cm}author(Author,[NodeAuthor$|$NodeBook],3),\\
\hspace*{1cm}title(Title,[NodeTitle$|$NodeBook],3). \\
\hline
\end{tabular}\\
}
}
In the evaluation, the goal will firstly
trigger the retrieval of the books for the author $\it 'Suciu'$. In
particular, it will retrieve the node numbers of {\it Suciu}'s books. It
is achieved due to the instantiation of the corresponding argument
in the goal. Afterward, it allows us the retrieval of $\it Suciu's$
book titles, ensuring that Suciu's book titles are the only
computed ones. 

The use of $\it author(Author, [NodeAuthor|NodeBook],3)$ is vital 
for the efficient retrieval of such titles, given that the node number has
been instantiated in this predicate in the first step. In this case,
the first used fact is $\it author('Suciu', [3,1,1],3)$ with the
node number $\it [3,1,1]$ and this node number is used for retrieving the fact $\it
title('Data~on~the~Web', [4,1,1],3)$.
Next, we show the (unique) computed answer by means of the evaluation as well
as the XML document represented by the goal instance:\\
{\fontsize{7pt}{9pt}\selectfont
{\it
\begin{tabular}{ll}
\hline
\hspace*{-3cm}\begin{tabular}{c}
{\it Title/'Data on the Web'}\\
{\it Review/Review'}\\
{\it Year/Year'}\\
{\it Node/[1, 1]}\\
\end{tabular} &
\hspace*{-7cm}\begin{tabular}{l}
$<$result$>$\\
\hspace*{1cm}$<$title$>$Data on the Web$<$/title$>$\\
$<$/result$>$
\end{tabular}\\
\hline
\end{tabular}\\
}
}
Let us remark that in the position of $\it year$ and $\it review$, which are not required in the {\it XPath} expression,
the goal returns variables (i.e. $\it Review'$, $\it Year'$). That is,
the evaluation does not use the facts for these elements. This is the main
effect of our specialization technique.

\subsubsection*{Example 4} 
Now, let us consider the {\it XPath} query
$\it /books/book[@year=2002 ~$ $\it and~title$ $\it =``Data~$ $on~the~Web"]/author$. In this case, the goal is $\it :-book$ $(\it booktype(Author,'Data$ $on$ $\it~the~Web',Review,$ $\it ['2002']),$ $\it Node,2)$, 
and the specialized schema rule is:\\
{\fontsize{7pt}{9pt}\selectfont
{\it
\begin{tabular}{l}
\hline
book(booktype(Author, Title, Review, [Year]),NodeBook,2):-\\
\hspace*{1cm}year(Year,NodeBook,3),\\
\hspace*{1cm}title(Title,[NodeTitle$|$NodeBook],3),\\
\hspace*{1cm}author(Author, [NodeAuthor$|$NodeBook],3).\\
\hline
\end{tabular}\\
}
}
In this specialized schema rule, we can see that the call to $\it review$ has been removed from the original schema rule, and the predicates
have been reordered with the aim of following the same order as the {\it XPath} expression. That is,
the boolean conditions are checked from left to right (firstly, $\it @year=2002$ and after
$\it title=``Data~on~the~Web"$), and finally, the authors are computed.
In other words, starting from the goal $\it book(booktype$ $(Author,'Data~on~the~Web',$ $\it Review,['2002']),Node,2)$,
firstly the retrieval of the books for the
year {\it 2002} is triggered. Afterward, the
retrieval of titles for this year (using the node number
instantiated in the previous step) is triggered; concretely the book titled $\it ``Data~on~the~$ $Web"$.
Finally, the authors of such books are retrieved
using node numbers instantiated in the previous steps.

In the case of an ``or" connective, that is, $\it /books/book[@year=2002 ~$ $\it or~title$ $\it =``Data~$ $on~the~Web"]/author$,
we would have two goals and patterns: $\it :-book(booktype$ $(Author,'Data~on~the~Web',$ $\it Review,[Year]),Node,2)$
and $\it :-book(booktype$ $(Author,$ $Title,$ $\it Review,['2002']),Node,2)$.

\subsubsection*{Example 5} 

Let us consider the {\it XPath} query $\it /books/book[@year=2002]/author$ $\it [name=``Serge"]$
with respect to the following XML document:\\
{\fontsize{7pt}{9pt}\selectfont
{\it
\begin{tabular}{l}
\hline
$<$books$>$\\
\hspace*{0,5cm}$<$book year=$``$2003$"$$>$\\
\hspace*{1cm}$<$author$>$Abiteboul$<$name$>$Serge$<$/name$>$$<$/author$>$\\
\hspace*{1cm}$<$title$>$Data on the Web$<$/title$>$\\
\hspace*{1cm}$<$review$>$A $<$em$>$fine$<$/em$>$ book.$<$/review$>$\\
\hspace*{0,5cm}$<$/book$>$\\
\hspace*{0,5cm}$<$book year=$``$2002$"$$>$\\
\hspace*{1cm}$<$author$>$Buneman $<$name$>$Peter$<$/name$>$$<$/author$>$\\
\hspace*{1cm}$<$title$>$XML in Scotland$<$/title$>$\\
\hspace*{0,5cm}$<$/book$>$\\
$<$/books$>$\\
\hline
\end{tabular}\\
}
}
In this case, we have two goals:
$\it :-book(booktype$ $\it (authortype(Unlabeled,$ $\it 'Serge',[]),$ $\it Title,$ $Review,$ $['2002']),$ $\it Node,2)$
and $\it :-book(booktype$ $\it (authortype(Unlabeled,$$\it'Ser\-ge',$ $[]),$ $\it Title,$ $['2002']),$ $\it Node,3)$. There are two goals 
because there are two weakly distinct records for the tag {\it book}: the first one has the subelement {\it review} but not the second one.

In this case, there are two patterns for the query, that is, $PT(/books/book[@year$ $=2002]/author$ $\it [name=``Serge"])=
\{booktype$ $\it (authortype(Unlabeled,$ $\it 'Serge',[]),$ $\it Title,$ $Review,['2002']),
booktype$ $\it (au$\-$thortype(Unlabeled,$ $\it 'Serge',[]),$ $Title, ['2002'])$ $\}$.
In addition, there are two type numbers, one for each pattern $TN(booktype$ $\it (authortype(Unlabeled,$ $\it 'Serge',[]),$ $\it Title,$ $Review,['2002']))=\{2\}$ and $TN(booktype$ $\it (au$\-$thortype(Unlabeled,$ $\it 'Serge',[]),$ $Title,$ $['2002']))=\{3\}$. 
Now, the specialized schema rules are:\\
{\fontsize{7pt}{9pt}\selectfont
{\it
\begin{tabular}{l}
\hline
book(booktype(Author, Title, Review, [Year]),NodeBook,2):-\\
\hspace*{1cm}author(Author, [NodeAuthor$|$NodeBook],3),\\
\hspace*{1cm}year(Year,NodeBook,3).\\
book(booktype(Author, Title,  [Year]),NodeBook,3):-\\
\hspace*{1cm}author(Author, [NodeAuthor$|$NodeBook],4),\\
\hspace*{1cm}year(Year,NodeBook,4).\\
\hline
\end{tabular}\\
}
}
{\fontsize{7pt}{9pt}\selectfont
{\it
\begin{tabular}{l}
\hline
author(authortype(Unlabeled,Name,[]),NodeAuthor,3):-\\
\hspace*{1cm}name(Name,[NodeName$|$NodeAuthor],4),\\
\hspace*{1cm}unlabeled(Unlabeled,[NodeUnlabeled$|$NodeAuthor],4).\\
author(authortype(Unlabaled,Name,[]),NodeAuthor,4):-\\
\hspace*{1cm}name(Name,[NodeName$|$NodeAuthor],5),\\
\hspace*{1cm}unlabeled(Unlabeled,[NodeUnlabeled$|$NodeAuthor],5).\\
\hline
\end{tabular}\\
}
}

\section{Theoretical Results}

In this section, we will prove the correctness of the proposed technique.
Our technique is correct in the sense that given a type and node numbered XML document ${\cal X}$,
the logic program ${\cal P}$ represented by ${\cal X}$,
and an {\it XPath} expression $xpathexpr$ then $subtree({\cal X},xpathexpr)=Doc(xpathexpr,{\cal P})$.
In other words, the subtree of an XML document defined by means of an {\it XPath} expression is the same
as the fragment of XML document build from the answers (w.r.t. the specialized schema rules) 
of the set of goal instances obtained from the same {\it XPath} expression. 

\begin{theorem}[Correctness]
Given a type and node numbered XML document ${\cal X}$, the logic program ${\cal P}$
represented by ${\cal X}$, and an {\it XPath}
expression $xpathexpr$, then
$subtree({\cal X},xpathexpr)=Doc(xpathexpr,{\cal P})$.\\
{\it Proof}\\
Let $xpathexpr$ be the {\it XPath} expression and let $xpathfree=FE(xpathexpr)$ be the free of equalities {\it XPath} expression
associated to $xpathexpr$.  
Now,  we have (1):
$$Doc(xpathexpr,{\cal P})=$$
$$Doc(Rules({\cal P}^{xpathexpr}) \cup \{tag(t,Node,r)\theta |  tag(t,Node,r) \in {\cal G}^{xpathexpr}\})$$  
by definition, where the $\theta$'s are answers w.r.t. ${\cal P}^{xpathexpr}$
and $t$ is a variable whenever $xpathexpr$ has no boolean conditions.
Moreover, (2): $${\cal P}^{xpathexpr} = Rules(Prog(subtree({\cal X},xpathfree))) \cup Facts({\cal P})$$
by definition. Let ${\cal F}$ be the set of facts used in the answers $\theta$ of $tag(t,Node,r)$:
$${\cal F}=_{def} \{ f\theta | f \in Facts({\cal P}), {\it f ~is~a~subgoal~of~tag(t,Node,r)~ in~the~branch~of~\theta},$$
$$tag(t,Node,r) \in {\cal G}^{xpathexpr}\}$$ 

Therefore, from (1) and (2), we have (3):
$$Doc(xpathexpr,{\cal P})=Doc(Rules({\cal P}^{xpathexpr}) \cup {\cal F})$$
Now, we have to prove that (4):
$$Doc(Rules({\cal P}^{xpathexpr}) \cup {\cal F}) = Doc(Rules(Prog(subtree({\cal X},xpathexpr)))$$ 
$$\cup Facts(Prog(subtree({\cal X},xpathexpr))))$$
 
To prove (4) we have to reason that (5):
$${\cal X'}=<tag'~ att_1=v_1,\dots,att_n=v_n,nodenumber=i,typenumber=k>$$
$$elem_1,\dots,elem_s </tag>$$ is a non terminal tagged subelement in $subtree({\cal X},xpathexpr)$  iff the schema rule 
$$tag'(tagtype'(\overline{Tag},[\overline{Att}]),Node,k):-C \in Rules(Prog(subtree({\cal X},xpathexpr)))$$
where $C$ is built from the tags of $elem_1,\dots,elem_s$ and $att_1,\dots,att_n$; and
${\cal X'}$ satisfies $expr_r$ where $xpathexpr=/expr_1 \dots /expr_r/ \dots /expr_m$;
and, in addition, (6):
$${\cal X'}=<tag'~ nodenumber=i,typenumber=k> elem </tag>$$
is a terminal tagged element in $subtree({\cal X},xpathexpr)$ iff $$tag'(elem,i,k) \in {\cal F}$$

(5) is obvious by definition.  Let us prove (6). We have to reason that if $f$ is a subgoal of $tag(t,Node,r)$
and $\theta$ is the answer of the branch including $f$ as subgoal, then if $f\theta$ is a fact 
we can map $f\theta$ into a terminal tagged subelement of $subtree({\cal X},xpathexpr)$.
It follows from the specialization of the schema rules of ${\cal P}$ and the choice of the patterns for 
$tag$.

Now, from (5) and (6) we can conclude (4) because if ${\cal X}'$ satisfies $expr_r$ then ${\cal X}'$ satisfies $FE(expr_r)$
by the definition of satisfiability, and therefore also:
$$tag'(tagtype'(\overline{Tag},[\overline{Att}]),Node,k):-C \in Rules(Prog(subtree({\cal X},xpathfree)))$$
and by (1):
$$Rules({\cal P}^{xpathexpr}) = Rules(Prog(subtree({\cal X},xpathfree)))$$

Now, from (3) and (4), and taking into account that:
$$subtree({\cal X},xpathexpr)=Doc(Rules(Prog(subtree({\cal X},xpathexpr)))$$ 
$$\cup Facts(Prog(subtree({\cal X},xpathexpr))))$$
which is trivially true, then we can conclude that:

$$subtree({\cal X},xpathexpr)=Doc(xpathexpr,{\cal P})$$

\end{theorem}
\section{Indexing}

In this section, we will describe how to index XML documents represented by means of a logic
program. In addition, we will show how to combine indexing and top-down evaluation.
The aim of the indexing is to improve the retrieval of facts from secondary memory
and therefore the execution of {\it XPath} queries. 

In summary, the storing model in our approach is as follows.

\begin{itemize}
\item We use \emph{main memory} for the storing of schema rules.
\item We use \emph{secondary memory  (i.e. files)} for the storing of facts.
\item We \emph{index} facts in secondary memory.
\item We have \emph{two kinds of indexes}: one for indexing \emph{predicate names},
and other one for indexing   \emph{group of facts}.
\end{itemize}

The use of main memory for storing the schema rules is justified due
to in most of cases the number of schema rules is small. The
use of secondary memory for storing facts is justified since
XML documents can be too big in order to be stored in main memory.

Fact indexing is justified for efficiency reasons. Firstly, our
approach requires to recover facts for a given predicate; in this
case we use the first kind of index.
Secondly, our approach requires to retrieve the elements
grouped in the same XML record (i.e. groups of facts refereed to the same XML record); 
in this case we use the second kind of index.

For instance, w.r.t. the running example, we  
 generate the following set of indexes:  

\begin{center}
{\fontsize{7pt}{9pt}\selectfont
{\it
\sf \hspace*{-4.25cm} first index \sf \hspace*{1cm} second index
\hspace*{0.75cm} group identifier \hspace*{1.2cm} facts\\
\begin{tabular}{llll}
\hline
\it author &
\hspace*{-5cm}
\begin{tabular}{l}
\it pos(1, 0).\\
\it pos(2, 0).\\
\it pos(3, 0).\\
\it pos(9, 8).\\
\end{tabular} &
\hspace*{-5cm}
$\it [1,1]$ & \hspace*{-6cm}\begin{tabular}{l}
\it (0) year('2003', [{\bf 1, 1}], 3).\\
\it (1) author('Abiteboul', [1, {\bf 1, 1}], 3).\\
\it (2) author('Buneman', [2, {\bf 1, 1}], 3).\\
\it (3) author('Suciu', [3, {\bf 1, 1}], 3).\\
\it (4) title('Data on the Web', [4, {\bf 1, 1}], 3).\\
\end{tabular}\\
\hline
\it em &
\hspace*{-5cm}\begin{tabular}{l}
\it \hspace*{0,5mm} pos(6,5).\\
\it \hspace*{0,5mm} pos(12,11).\\
\end{tabular}  &
\hspace*{-5cm}
$\it [5,1,1]$ & \hspace*{-6cm}\begin{tabular}{l}
\it \hspace*{-4mm} (5) unlabeled('A ',  [1, {\bf 5, 1, 1}], 4).\\
\it \hspace*{-4mm} (6) em(fine, [2, {\bf 5, 1, 1}], 4).\\
\it \hspace*{-4mm} (7) unlabeled(' book.', [3, {\bf 5, 1, 1}], 4).\\
\end{tabular}\\
\hline
\it title &
\hspace*{-5cm}\begin{tabular}{l}
\it \hspace*{1mm} pos(4, 0).\\
\it \hspace*{1mm} pos(10, 8).\\
\end{tabular} &
\hspace*{-5cm} $\it [2,1]$ & \hspace*{-6cm}\begin{tabular}{l}
\it \hspace*{1mm} (8) year('2002', [{\bf 2, 1}], 3).\\
\it \hspace*{1mm} (9) author('Buneman', [1, {\bf 2, 1}], 3).\\
\it \hspace*{1mm} (10) title('XML in Scotland', [2, {\bf 2, 1}], 3).\\
\end{tabular}\\
\hline
\it unlabeled &
\hspace*{-5cm}\begin{tabular}{l}
\it \hspace*{1mm} pos(5, 5).\\
\it \hspace*{1mm} pos(7, 5).\\
\it \hspace*{1mm} pos(11, 11).\\
\it \hspace*{1mm} pos(13, 11).\\
\end{tabular} &
\hspace*{-5cm}
$\it [1,3,2,1]$ & \hspace*{-6cm}\begin{tabular}{l}
\it \hspace*{1mm} (11) unlabeled('The ', [1, {\bf 1, 3, 2, 1}], 6).\\
\it \hspace*{1mm} (12) em(best, [2, {\bf 1, 3, 2, 1}], 6).\\
\it \hspace*{1mm} (13) unlabeled(' ever!', [3, {\bf 1, 3, 2, 1}], 6).\\
\end{tabular}\\
\hline
\it year &
\hspace*{-5cm}\begin{tabular}{l}
\it pos(0, 0).\\
\it pos(8, 8).\\
\end{tabular} &
\hspace*{-1cm}
 ~ & \hspace*{-2cm}\begin{tabular}{l}
~
\end{tabular}\\
\hline
\end{tabular}
}
}
\end{center}

The first index allows the retrieval of facts by means of the predicate name: 
$\it author$, $\it year$, and so on. 
Therefore, the \emph{first index key} is the \emph{name of the predicate} and the \emph{first index value} is the \emph{set of relative
positions in the file} of the facts for the predicate.

The second index allows to recover the relative position in the file of the group in which a fact is included.
Therefore the \emph{second index key} is the \emph{relative position of the fact in the file} and
the \emph{second index value} is the \emph{relative position in the file of the group  in which the fact is included}.

With this aim the first index stores for each predicate name annotations of the form $\it pos(n,m)$,
in which $\it n$ denotes the relative position in the file of a fact for the predicate and $\it m$ the
relative position in the file of the group of this fact (therefore the second index is a secondary index).

For instance, $\it author$ facts are stored in positions $\it 1$, $\it 2$, $\it 3$ and $\it 9$, given by
the annotation $\it pos({\bf 1},0)$, $\it pos({\bf 2},0)$, $\it pos({\bf 3},0)$, $\it pos({\bf 9},8)$, and the group of
each author, that is, the XML record in which the author is included, starts at positions $\it 0$, $\it 0$, $\it 0$
and $\it 8$, respectively, given by the annotations  $\it pos(1,{\bf 0})$, $\it pos(2,{\bf 0})$,
$\it pos(3,{\bf 0})$, $\it pos(9,{\bf 8})$.  Each ``group of facts" shares the \emph{node number
of the record}, which can be considered as the \emph{identifier of the group}.

For instance, w.r.t. the running example,
the first group can be identified by $\it [1,1]$, and contains facts numbered as
$\it [1,1]$, $\it [1,1,1]$, $\it [2,1,1]$, $\it [3,1,1]$ and $\it [4,1,1]$. 
The second group is $\it [5,1,1]$, and so on.
The reason for this grouping criteria is that each group of facts
will be retrieved by means of the same schema rule. For instance, in the running example, the schema
rule:
\begin{center}
\noindent {\fontsize{7pt}{9pt}\selectfont
{\it
\begin{tabular}{l}
\hline\\[-2mm]
book(booktype(Author, Title, Review, [Year]), NodeBook ,2) :- \\
\hspace*{1cm}author(Author, [NodeAuthor$|$NodeBook],3),\\
\hspace*{1cm}title(Title, [NodeTitle$|$NodeBook],3),\\
\hspace*{1cm}review(Review, [NodeReview$|$NodeBook],3),\\
\hspace*{1cm}year(Year, NodeBook,3).\\
\hline
\end{tabular}
}
}
\end{center}
will retrieve the groups of facts $[1,1]$ and $[2,1]$.

Now, we will explain how the indexing technique
is combined with the top-down evaluation of the goals. 
For instance, let us suppose the following {\it XPath} query:
$\it /books/book[@year=2002~and~author=``Buneman"]/review$ w.r.t. the running example.
Now, the specialized schema rules and facts used in the evaluation are:

\begin{center}
\noindent {\fontsize{7pt}{9pt}\selectfont
{\it
\begin{tabular}{l}
\hline
\begin{tabular}{l}
(a) book(booktype(Author, Title, Review, [Year]), NodeBook ,2) :-        \\
\hspace*{1cm}year(Year, NodeBook,3),\\
\hspace*{1cm}author(Author, [NodeAuthor$|$NodeBook],3),\\
\hspace*{1cm}review(Review, [NodeReview$|$NodeBook],3).\\
(b) review(reviewtype(Unlabeled,Em,[]),NodeReview,3):-\\
\hspace*{1cm}unlabeled(Unlabeled,[NodeUnlabeled$|$NodeReview],4),\\
\hspace*{1cm}em(Em,[NodeEm$|$NodeReview],4).\\
(c) review(reviewtype(Em,[]),NodeReview,3):-\\
\hspace*{1cm}em(Em,[NodeEm$|$NodeReview],5).\\
(d) em(emtype(Unlabeled,Em,[]),NodeEms,5) :-        \\
\hspace*{1cm}unlabeled(Unlabeled,[NodeUnlabeled$|$NodeEms],6),\\
\hspace*{1cm}em(Em, [NodeEm$|$NodeEms],6).\\
\end{tabular}
\end{tabular}
}
}
\end{center}
\begin{center}
\noindent {\fontsize{7pt}{9pt}\selectfont
{\it
\begin{tabular}{l}
 \begin{tabular}{l}
(0) year('2003', [1, 1], 3).\\
(1) author('Abiteboul', [1, 1, 1], 3).\\
(2) author('Buneman', [2,1, 1], 3).\\
(3) author('Suciu', [3,1,1], 3).\\
(4) title('Data on the Web', [4, 1, 1], 3).\\
(5) unlabeled('A', [1, 5, 1, 1], 4).\\
(6) em('fine', [2, 5, 1, 1], 4).\\
(7) unlabeled('book.', [3, 5, 1, 1], 4).\\
(8) year('2002', [2, 1], 3).\\
(9) author('Buneman', [1, 2, 1], 3).\\
(10) title('XML in Scotland', [2, 2, 1], 3).\\
(11) unlabeled('The', [1, 1, 3, 2, 1], 6).\\
(12) em('best', [2, 1, 3, 2, 1], 6).\\
(13) unlabeled('ever!', [3, 1, 3, 2, 1], 6).\\
\end{tabular}\\
\hline
\end{tabular}
}
}
\end{center}
 
The combination of indexing and top-down evaluation can be summarized as follows.
In general,  the evaluation will generate (sub)goals which have the form:
$\it tag(\_,[Var_1,$ $\it \dots,Var_n,N_1,\dots,N_m],\_)$, where $\it tag$ is
a tag of the XML document.
The second argument of such (sub)goals 
is a list of the form $\it [Var_1,\dots,Var_n,$ $\it N_1,$ $\dots,N_m]$
representing a \emph{partially instantiated node number}, in which
$\it Var_1,$ $\dots,Var_n$ are variables and $\it N_1,\dots,N_m$ are natural numbers.
There is a particular case of goals of the form $\it tag(\_,Var,\_)$, in which there is a variable
in the second argument instead of a list. This particular case corresponds with the main goal.

In addition, each time a fact is recovered, the system stores, together with the
identifier of its group, the relative position in the file of its group.
For instance, w.r.t. the running example, whenever $\it author('Buneman', [2,1, 1], 3)$
is recovered, the system stores that the group $\it [1,1]$ is at position $\it 0$ in the file.

Now, the index accessing can be summarized as follows.
Each time a subgoal $\it tag(\_,$ $\it [Var_1,$ $\it \dots,Var_n,$ $\it N_1,\dots,N_m],\_)$ is called and
does not unify with an schema rule then:
\begin{itemize}
\item[(a)] Whenever $\it [Var_2,\dots,Var_n,$ $\it N_1,\dots,N_m]$ \emph{matches to a previously
stored group identifier}, the system uses the relative position of the matched group for the retrieval of facts for $\it tag$.
Therefore the second index is used for the retrieval of the facts.
\item[(b)] Whenever the stored group identifiers \emph{do not match} to $\it [Var_2,$ $\dots,Var_n,$ $\it N_1,\dots,$ $N_m]$,
the system uses the first index for the retrieval of the elements of $\it tag$.
\end{itemize}
In the case of the main goal $\it tag(\_,Var,\_)$, the first index will be ever used.

Now, we show the \emph{trace of the execution} of the
{\it XPath} query $\it /books/book[@year=2002~and~author=``Buneman"]/review$ with respect to the above indexing structure.

\rule{12.5cm}{.5mm}

{\scriptsize
\begin{enumerate}
\item call of $\it book(booktype(Buneman, \_G12073, \_G12074, [2002]), \_G12078, 2)$ (Rule a)
\item call of $\it year(2002, \_G12128, 3)$ (Rule a)
\item first index accessing to position $\it 0$ due to $\it year(2002, \_G12128, 3)$;
recovering  $\it year(2003, [1, 1], 3)$; {\bf fail}.
\item first index accessing to position $\it 8$ due to $\it year(2002, \_G12128, 3)$;
recovering  $\it year(2002, [2, 1], 3)$; storing that the position of group $\it [2,1]$ is $\it 8$; {\bf success}.
\item call of $\it author($ $Buneman, [\_G12100, 2, 1], 3)$ (Rule a)
\item second index accessing to position $\it 8$ due to the position of group $\it [2,1]$ is $\it 8$;
 recovering $\it author($ $Buneman, $ $\it [1, 2, 1], 3)$; {\bf success}
\item call of $\it review(\_G12151, [\_G12148, 2, 1], 3)$ (Rule a)
\item call of $\it unlabeled(\_G12190, [\_G12187, \_G12212, 2, 1], 4)$ (Rule b)
\item first index accessing to position $\it 11$
due to $\it unlabeled(\_G12243, [\_G12240, \_G12265, \_G12268, 2, 1],$ $ 6)$;
recovering $\it unlabeled$ $\it (The , [1, 1, 3, 2, 1], 6)$; storing that the position
of group $\it [1,3,2,1]$ is $\it 11$; {\bf success}.
\item first index accessing to position $\it 13$ due to
$\it unlabeled(\_G12243, [\_G12240, \_G12265, \_G12268, 2, 1],$ $ 6)$;
recovering $\it unlabeled$ $\it (ever!, [3, 1, 3, 2, 1], 6)$; storing that position
of group $\it [1,3,2,1]$ is $\it 11$; {\bf success}
\item call of  $\it em(\_G12261, [\_G12258, \_G12283, \_G12286, 2, 1], 6)$ (Rule c)
\item second index accessing to position $\it 11$ due to
$\it em(\_G12261, [\_G12258, \_G12283, \_G12286, 2, 1], 6)$ and that position of
group $\it [1,3,2,1]$ is $\it 11$; recovering  $\it em(best, [2, 1, 3, 2, 1], 6)$; {\bf success}
\item $\it em(emtype(The , best, [\,]), [1, 3, 2, 1], 5)$ {\bf success}
\item $\it em(emtype(ever!, best, [\,]), [1, 3, 2, 1], 5)$ {\bf success}
\item $\it review(reviewtype(emtype(The , best, [\,]), [\,]), [3, 2, 1], 3)$ {\bf success}
\item $\it review(reviewtype(emtype( ever!, best, [\,]), [\,]), [3, 2, 1], 3)$ {\bf success}
\item $\it book(booktype(Buneman, \_G12316, reviewtype(emtype(The , best, [\,]), [\,]), [2002]), [2, 1], 2)$ {\bf success}
\item $\it book(booktype(Buneman, \_G12316, reviewtype(emtype( ever!, best, [\,]), [\,]), [2002]), [2, 1], 2)$ {\bf success}
\end{enumerate}
}
\rule{12.5cm}{.5mm}

\section{Prototype}
Now, we will show our prototype, named {\it XIndalog}. 
This prototype implements the technique presented in this paper.
In addition, we have implemented a rich set of {\it XPath} queries including
{\it XPath} constructions like  ``//", ``/../".``*", etc.  The prototype has
been developed under {\it SWI-Prolog} \cite{swi} and hosted in a web site
 at \url{http://indalog.ual.es/Xindalog}. This
web site has been developed  by using a {\it CGI} ({\it Common
Gateway Interface}) application, in order to link the web site with the prototype.
From the main page of the prototype (see Figure 2),
we can access to a basic description of {\it XIndalog}, {\it XML},
 {\it XPath}, as well as the {\it demo}.\\

\noindent\begin{tabular}{cc}
{\small {\bf Fig. 2.} \url{http://indalog.ual.es/Xindalog}} &
{\small {\bf Fig. 3.} Top-Down demo}\\
\includegraphics[width=6cm,height=5cm]{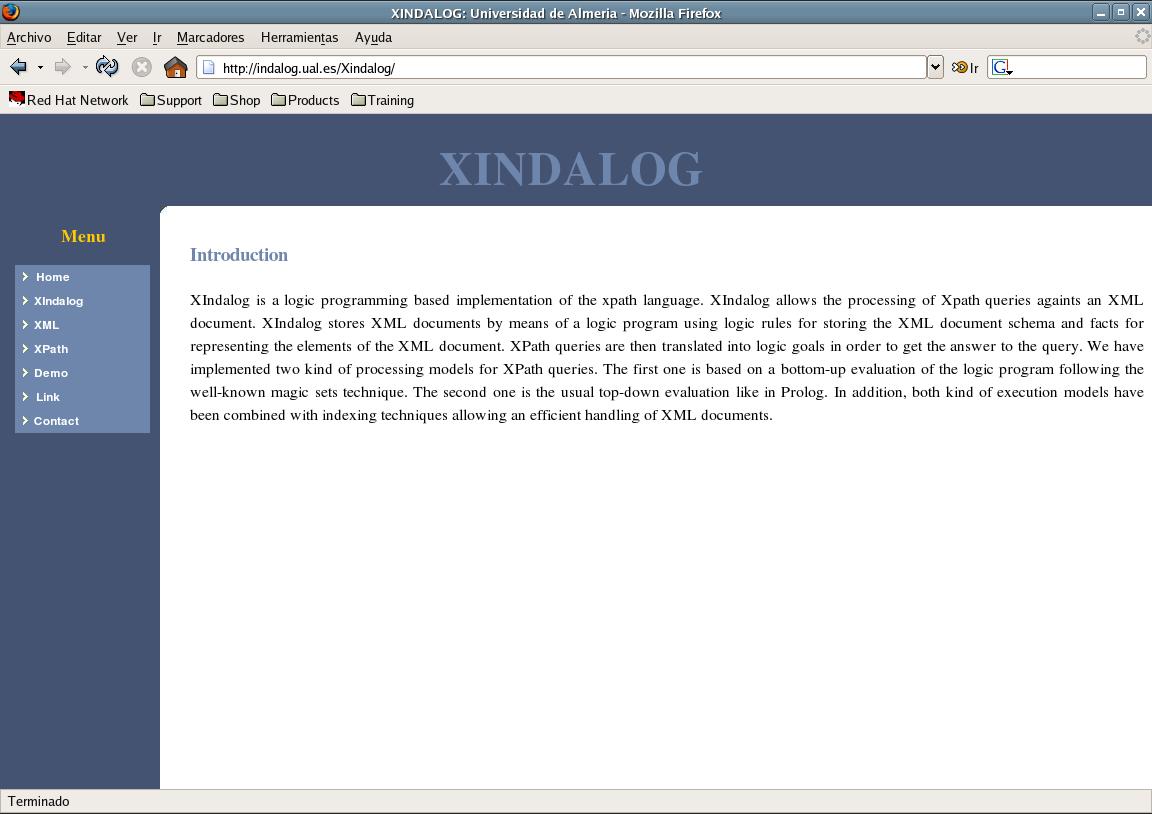}
&
\includegraphics[width=6cm,height=5cm]{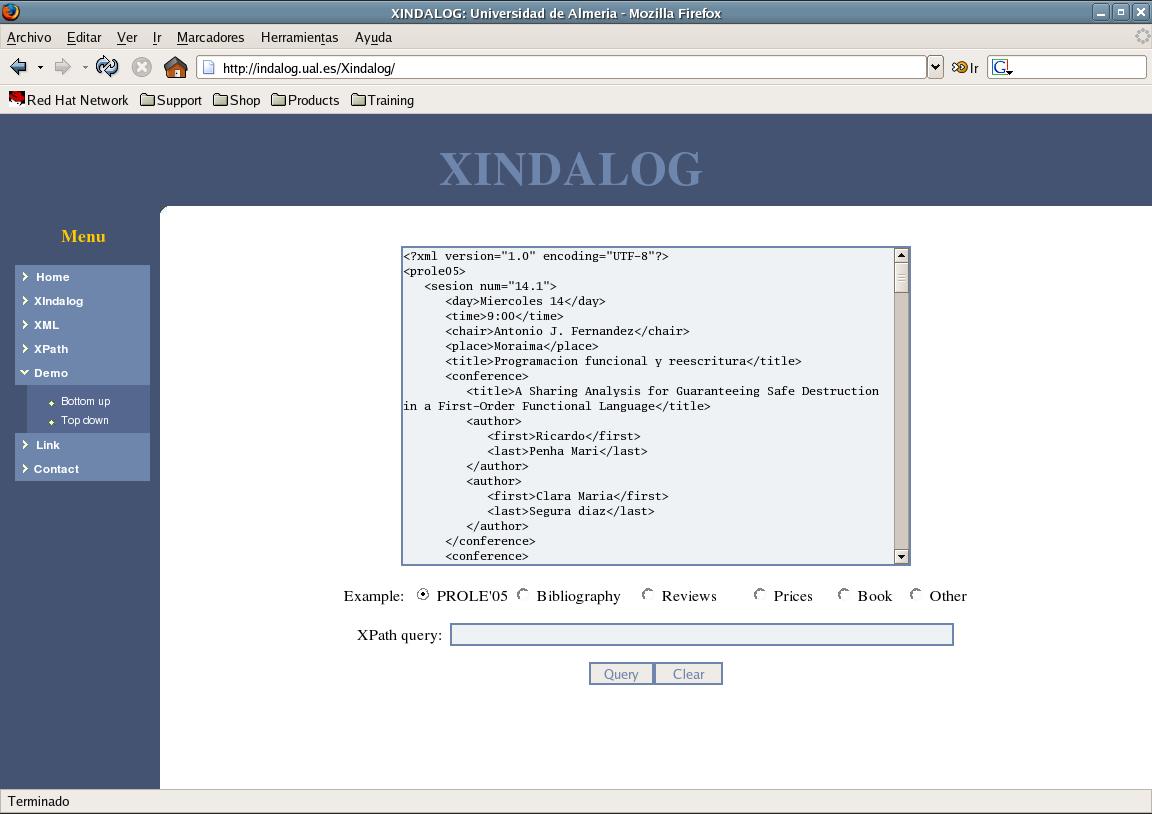}\\
\end{tabular}

We have implemented two releases of the
prototype: a top-down and bottom-up release (details about the later can be
found in \cite{jesusjucs}). In the web site, there are some built-in examples 
which can be tested and new examples can also be typed.\\

\noindent\begin{tabular}{cc}
{\small {\bf Fig. 4.} Query example} &
{\small {\bf Fig. 5.} Query result}\\
\includegraphics[width=6cm,height=5cm]{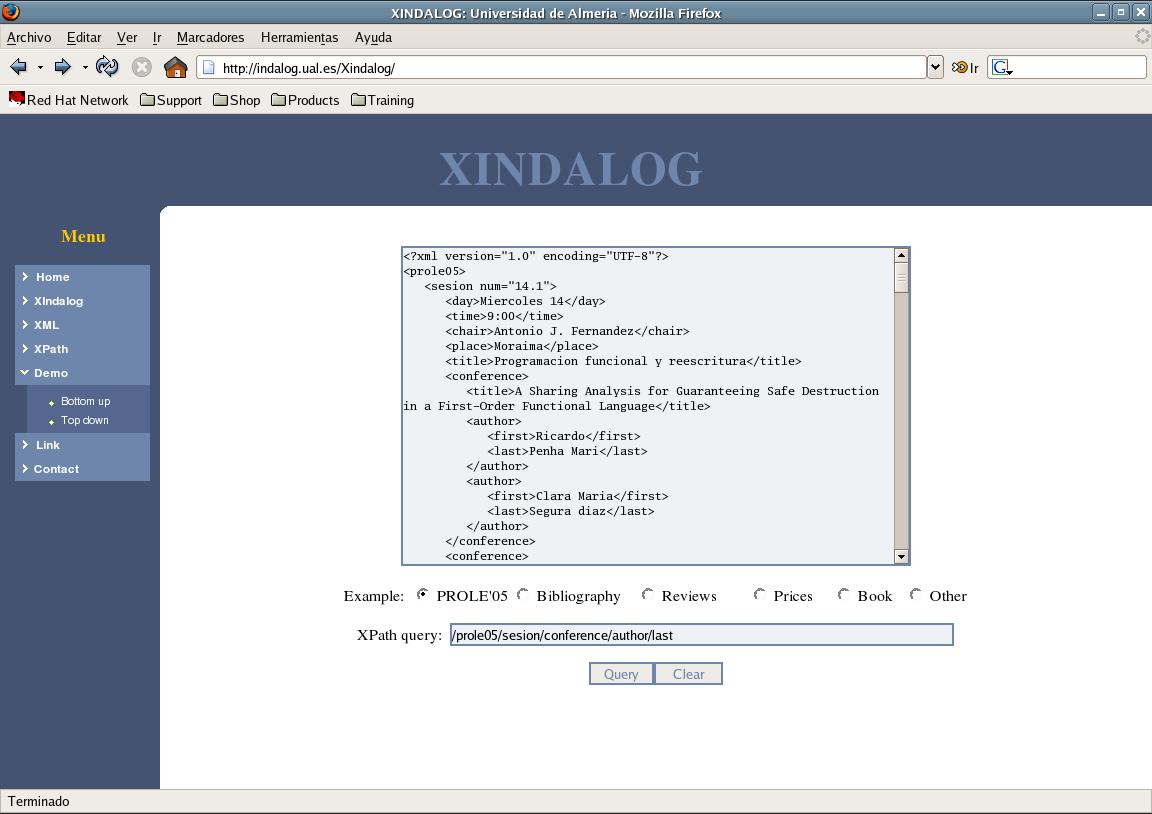}
&
\includegraphics[width=6cm,height=5cm]{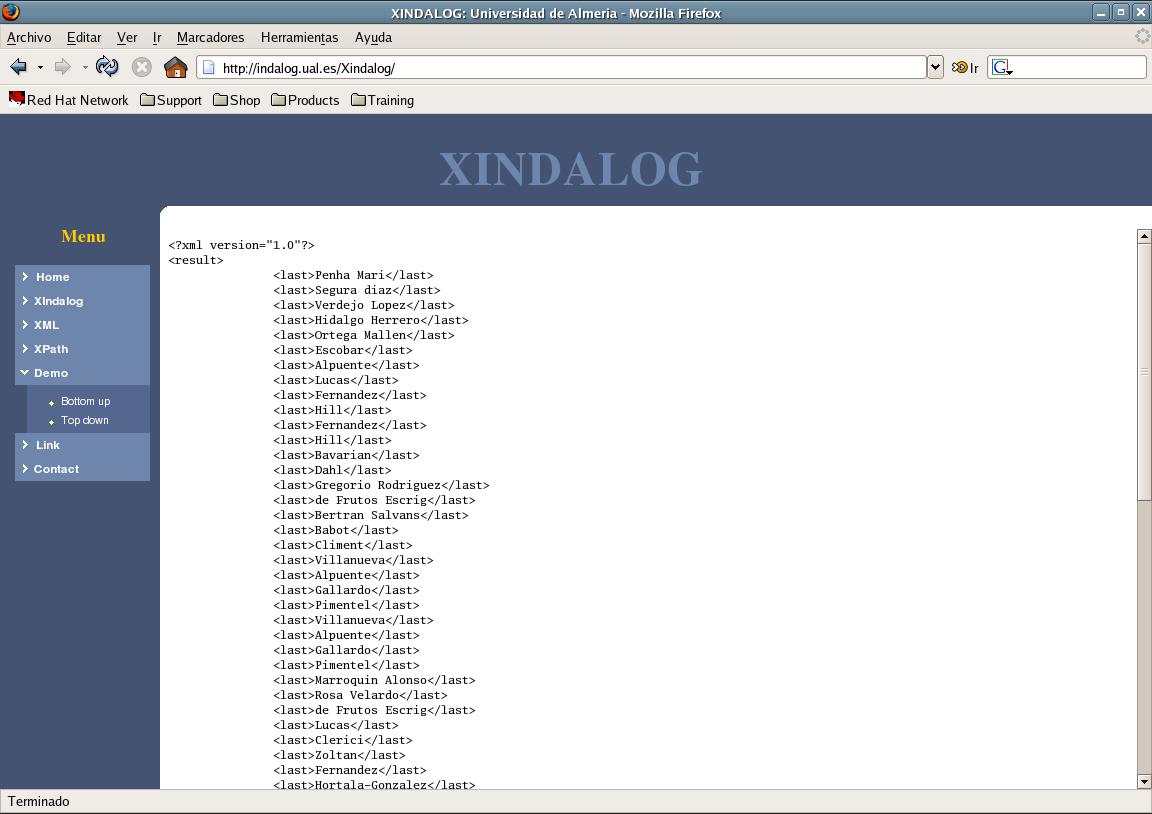}\\
\end{tabular}

\begin{table}[!t]
\label{table:xmldocumentone}
\caption{A small {\it XML} document}

\begin{center}
{\scriptsize \it
\begin{tabular}{l}
\hline
$<$books year=``2006"$>$\\
\hspace*{5mm}{\it A book collection}\\
\hspace*{5mm}$<$book$>${\it empty}$<$/book$>$\\
\hspace*{5mm}$<$book year=``2003" pages=``984"$>$\\
\hspace*{10mm}{\it The first book}\\
\hspace*{10mm}$<$author english=``yes" spanish=``yes"$>$\\
\hspace*{15mm}{\it Benz}\\
\hspace*{15mm}$<$name$>${\it Brian}$<$/name$>$\\
\hspace*{10mm}$<$/author$>$\\
\hspace*{10mm}$<$author$>${\it John Durant}$<$/author$>$\\
\hspace*{10mm}$<$author$>${\it John Durant}$<$/author$>$\\
\hspace*{10mm}$<$title$>${\it XML Programming Bible}$<$/title$>$\\
\hspace*{10mm}$<$review$>${\it Good}$<$/review$>$\\
\hspace*{5mm}$<$/book$>$\\
\hspace*{5mm}$<$book year=``2002"$>$\\
\hspace*{10mm}{\it The second book}\\
\hspace*{10mm}$<$author$>${\it Dino Esposito}$<$/author$>$\\
\hspace*{10mm}$<$title$>${\it Applied XML Programming for Microsoft .NET}$<$/title$>$\\
\hspace*{10mm}$<$review$>${\it Good}$<$/review$>$\\
\hspace*{5mm}$<$/book$>$\\
\hspace*{5mm}$<$book$>$\\
\hspace*{5mm}{\it The third book}\\
\hspace*{10mm}$<$author$>${\it Apt, Krzystof R.}$<$/author$>$\\
\hspace*{10mm}$<$title$>${\it The Logic Programming Paradigm and Prolog}$<$/title$>$\\
\hspace*{10mm}$<$review$>${\it Very good}$<$/review$>$\\
\hspace*{5mm}$<$/book$>$\\
\hspace*{5mm}$<$book year=``1994" pages=``560"$>$\\
\hspace*{10mm}{\it The fourth book}\\
\hspace*{10mm}$<$author english=``yes" spanish=``no"$>$\\
\hspace*{15mm}{\it Leon Sterling}\\
\hspace*{10mm}$<$/author$>$\\
\hspace*{10mm}$<$author$>${\it Ehud Shapiro}$<$/author$>$\\
\hspace*{10mm}$<$title$>${\it The Art of Prolog}$<$/title$>$\\
\hspace*{10mm}$<$review$>${\it Very good}$<$/review$>$\\
\hspace*{5mm}$<$/book$>$\\
\hspace*{5mm}$<$book2 year=``2001"$>$\\
\hspace*{10mm}{\it The fifth book}\\
\hspace*{10mm}$<$author english=``yes"$>$\\
\hspace*{15mm}{\it Elliotte Rusty Harold}\\
\hspace*{10mm}$<$/author$>$\\
\hspace*{10mm}$<$title$>${\it XML Bible}$<$/title$>$\\
\hspace*{10mm}$<$review2$>${\it Good}$<$/review2$>$\\
\hspace*{5mm}$<$/book2$>$\\
\hspace*{5mm}$<$book year=``2003" pages=``984"$>$\\
\hspace*{10mm}{\it The first book}\\
\hspace*{10mm}$<$author english=``yes" spanish=``yes"$>$\\
\hspace*{15mm}{\it Benz}\\
\hspace*{15mm}$<$name2$>${\it Brian}$<$/name2$>$\\
\hspace*{15mm}$<$firstname$>$\\
\hspace*{15mm}{\it Brian}\\
\hspace*{15mm}$<$lastname$>${\it Benz}$<$/lastname$>$\\
\hspace*{15mm}$<$others$>${\it no more}$<$/others$>$\\
\hspace*{15mm}$<$/firstname$>$\\
\hspace*{10mm}$<$/author$>$\\
\hspace*{10mm}$<$author$>${\it John Durant}$<$/author$>$\\
\hspace*{10mm}$<$author$>${\it John Durant}$<$/author$>$\\
\hspace*{10mm}$<$title$>${\it XML Programming Bible}$<$/title$>$\\
\hspace*{10mm}$<$review$>${\it Very good 2}$<$/review$>$\\
\hspace*{5mm}$<$/book$>$\\
$<$/books$>$\\
\hline
\end{tabular}
}
\end{center}

\end{table}

\subsection{Benchmarks}

We have tested our prototype by means of not enough structured XML documents and by means
of XML documents of big size.
Firstly, we have tested our prototype with a small but not enough structure XML document, shown in Table 1.
Now and w.r.t. this document, we have considered the following set of {\it XPath} queries.\\[-3mm]

\begin{center}
{\fontsize{7pt}{8pt}\selectfont \noindent \begin{tabular}{ll|l}
& \qquad \qquad {\bf XPath Query} & \qquad \qquad \qquad {\bf Meaning}\\
\hline
$\odot$ & /books/book[@year and @pages]/* & To obtain the books which have \\
&& publishing year and number of pages\\
$\odot$ & /books/book/author/@* & To obtain all the attributes of the authors\\
$\odot$ & //book & To obtain all the books\\
&& included in the {\it XML} document\\
\hline
\end{tabular}
}
\end{center}

\begin{center}
{\fontsize{7pt}{8pt}\selectfont \noindent \begin{tabular}{ll|l}
& \qquad \qquad {\bf XPath Query} & \qquad \qquad {\bf Meaning}\\
\hline
$\odot$ & //book[review=``Very good"]/author & To obtain all the authors\\
&& of books with a very good review\\
$\odot$ & //@year & To obtain all the years occurring in \\
 && the XML document\\
$\odot$ & /books/*/author & To obtain all the authors \\
&& inside book records\\
\hline
\end{tabular}
}
\end{center}

\begin{center}
{\fontsize{7pt}{8pt}\selectfont \noindent \begin{tabular}{ll|l}
& \quad \qquad \qquad {\bf XPath Query} & \qquad \qquad \qquad {\bf Meaning}\\
\hline
$\odot$ & /books/book[review=``Good"]/ & To obtain
 all the author information  \\
& author[name=``John Durant"] & whose name is John Durant and the review is good\\
$\odot$ & /books[book=``The first book"]/book  & To obtain the books\\
& [@year=2003 and review=``Good"] &
  of the year 2003 and good review \\
& /author[name=``Benz"]/../.. & whose author is Benz \\
$\odot$ & /books/book/text() & To obtain the books with textual information\\
$\odot$ & /books/book[author/name]/title & To obtain the book titles  whenever\\
&& the books have author name\\
$\odot$ & /books/(book $\vert$ book2)/(review2 $\vert$ review) &  To obtain the reviews\\
&& of the two kinds of books\\
$\odot$ & /books/book/(author $\vert$ title) & To obtain the book authors and titles\\
$\odot$ & /books/(book $\vert$ book2)//text() & To obtain the textual information\\
&& from the two kinds of books\\
$\odot$ & //@* & To obtain all the attributes of the document\\
$\odot$ & /*/*/title & To obtain the titles that are at 3rd level\\
$\odot$ & /*/*//* & To obtain all the elements and their nested\\
&&  from the 3rd level\\
$\odot$ & /*/book2/* & To obtain all information from book2 at 2nd level\\
$\odot$ & //*//author/.. & To obtain the records containing \\
&& author information from the 1st level\\
\hline
\end{tabular}
}
\end{center}

Secondly, we have tested our prototype with XML documents of big size in order
to get benchmarks, considering the following file sizes:
\begin{itemize}
\item {\it 64KB}; 516 elements were included into the file;
\item {\it 128KB}; 1032 elements were included into the file;
\item {\it 256KB}; 2064 elements were included into the file;
\item {\it 512KB}; 4128 elements were included into the file; and finally,
\item {\it 1024KB}; 8256 elements.
\end{itemize}
For each file size, we have computed the following answer times:
\begin{itemize}
\item {\it Translation time};\\
It represents the time needed for translating
 a {\it XML} document into {\it Prolog} facts and rules;
\item {\it Evaluation time};\\
 It represents the time of the top-down evaluation of the
 (specialized) program w.r.t. an {\it XPath} query;
\item {\it Browsing time};\\
 It represents the time needed for
 formatting and browsing the query result.
\end{itemize}
Next, we will show three {\it XPath} queries with their corresponding times for
 each considered file size.\\

\noindent \underline{{\it XPath} Query}: {\it /books}

\begin{center}
\noindent\begin{tabular}{c|c|c|c|c}
{\it File size} & {\it Translation} & {\it Evaluation} & {\it Browsing} & {\it Total time}\\
\hline
64KB & 1,063sg & 2,062sg & 0,063sg & 3,188sg\\
128KB & 3,375sg & 7,717sg & 0,125sg & 11,2171sg\\
256KB & 11,860sg & 31,296sg & 0,312sg & 43,4681sg\\
512KB & 42,812sg & 2min 11,110sg & 0,578sg & 2min 54,500sg\\
\hline
\end{tabular}
\end{center}

\noindent \underline{{\it XPath} Query}: {\it /books/book/title}

\begin{center}
\noindent\begin{tabular}{c|c|c|c|c}
{\it File size} & {\it Translation} & {\it Evaluation} & {\it Browsing} & {\it Total time}\\
\hline
64KB & 1,030sg & 0,204sg & 0,030sg & 1,264sg\\
128KB & 3,343sg & 0,673sg & 0,047sg & 4,063sg\\
256KB & 11,546sg & 2,484sg & 0,048sg & 14,078sg\\
512KB & 42,813sg & 9,562sg & 0,188sg & 52,563sg\\
\hline
\end{tabular}
\end{center}

\noindent \underline{{\it XPath} Query}: {\it /books/book[review=``very good"]/title}

\begin{center}
\noindent\begin{tabular}{c|c|c|c|c}
{\it File size} & {\it Translation} & {\it Evaluation} & {\it Browsing} & {\it Total time}\\
\hline
64KB & 1,046sg & 0,032sg & 0,0sg & 1,078sg\\
128KB & 3,359sg & 0,063sg & 0,0sg & 3,422sg\\
256KB & 11,579sg & 0,108sg & 0,0sg & 11,687sg\\
512KB & 42,796sg & 0,188sg & 0,0sg & 42,984sg\\
\hline
\end{tabular}
\end{center}
The following tables show the benchmarks of the query {\it
/books/book[review=``good"] /title} with and without our program
specialization technique. From these tables, we can conclude that \emph{our
specialization technique significantly improves the
answer times}.\\

\noindent \underline{{\it XPath} Query}: {\it /books/book[review=``good"]/title}\\[2mm]
\noindent {\it \underline{Without Program Specialization}}
\begin{center}
\noindent\begin{tabular}{c|c|c|c|c}
{\it File size} & {\it Translation} & {\it Evaluation} & {\it Browsing} & {\it Total time}\\
\hline
64KB & 0,750sg & 1,562sg & 0,046sg & 2,358sg\\
128KB & 2,095sg & 5,202sg & 0,095sg & 7,392sg\\
256KB & 6,579sg & 19,407sg & 0,187sg & 26,173sg\\
512KB & 22,530sg & 1min 21,172sg & 0,500sg & 1min 44,202sg\\
1024KB & 1min 22sg & 5min 32,843sg & 0,921sg &  6min 55,764sg\\
\hline
\end{tabular}
\end{center}
{\it \underline{With Program Specialization}}\\
\begin{center}
\noindent\begin{tabular}{c|c|c|c|c}
{\it File size} & {\it Translation} & {\it Evaluation} & {\it Browsing} & {\it Total time}\\
\hline
64KB & 0,750sg & 0,172sg & 0,015sg & 0,937sg\\
128KB & 2,079sg & 0,546sg & 0,0165sg & 2,641sg\\
256KB & 6,484sg & 2sg & 0,048sg & 8,532sg\\
512KB & 22,298sg & 7,656sg & 0,094sg & 30,048sg\\
1024KB & 1min 21,546sg & 30,296sg & 0,188sg &  1min 52,030sg\\
\hline
\end{tabular}
\end{center}

\section{Conclusions and Future Work}

In this paper, we have presented how to represent and index XML documents by means
of logic programming. Moreover, we have studied how to specialize a logic program, and how to generate 
goals in order to solve {\it XPath} queries.
We have described how to use the indexing of the XML documents 
in order to obtain a more efficient top-down evaluation and query solving.
Finally, we have shown benchmarks of our prototype developed 
with the proposed technique. Our approach opens two promising research lines.

\begin{itemize}

\item The first one, the extension of {\it XPath} to
a more powerful query language such as {\it XQuery}, that is, the study of the implementation of {\it XQuery} in logic programming.
 
 We have developed an extension to {\it XQuery} in a recent paper \cite{jesuswlp}, which uses as basis
 the specialization technique studied here for {\it XPath} queries. {\it XQuery} enriches our proposal since 
 in {\it XQuery} the queries can involve more than one XML document. In addition, {\it XQuery} allows us
 to express more complex queries w.r.t. a sole document. Now, we are developing the implementation
 of our new proposal. 

\item The second one, the use of logic programming as inference engine for the so-called {\it ``Semantic Web"}, by
introducing {\it RDF} documents or {\it OWL} specifications. In this line we are interested in the representation in
logic programming of ontologies. 

There are some recent works \cite{Wolz,Grosof,HorrocksOWL} interested in the identification of the 
intersection of logic programming and the so-called {\it Description Logic (DL)} \cite{DL}, the basis of most ontology languages. 
The quoted proposals translate restricted forms of ontologies (i.e. restricted forms of {\it OWL} and therefore fragments of {\it DL}) into logic programming. 
 Our work can be integrated in this framework
by combining our logic programming based transformation of XML documents and  the transformation
of ontologies into logic programming.  

The interest of such integration is to provide semantic information about XML documents, the use 
of such semantic information in order to inferring new information, and thus to improve the answers to {\it XPath} and {\it XQuery} queries.
\end{itemize}

\end{document}